\documentclass[twocolumn]{aastex631}
\usepackage{lineno}

\newcommand\aastex{AAS\TeX}

\usepackage{xcolor}
\usepackage{wrapfig}
\usepackage{url}
\usepackage{multirow}
\usepackage{booktabs} 
\usepackage[T1]{fontenc} 
\usepackage[utf8]{inputenc}  
\usepackage[graphicx]{realboxes} 

\usepackage{tipa}

\shorttitle{\aastex\ Rocky Planet Compositions Tend to be Earth-like}
\shortauthors{Brinkman et al.}

\begin{document}
\newcommand{\PSUAA}{Department of Astronomy \& Astrophysics, 525 Davey Laboratory, The Pennsylvania State University, University Park, PA, 16802, USA}
\newcommand{\PSUCEHW}{Center for Exoplanets \& Habitable Worlds, 525 Davey Laboratory, The Pennsylvania State University, University Park, PA 16802, USA}
\newcommand{\PSETI}{Penn State Extraterrestrial Intelligence Center, 525 Davey Laboratory, The Pennsylvania State University, University Park, PA 16802, USA}
\newcommand{\UA}{Steward Observatory, The University of Arizona, 933 N.\ Cherry Ave, Tucson, AZ 85721, USA}
\newcommand{\Caltech}{Department of Astronomy, California Institute of Technology, Pasadena, CA 91125, USA}
\newcommand{\JHU}{Department of Physics \& Astronomy, Bloomberg Center, Johns Hopkins University, Baltimore, MD 21218, USA}
\newcommand{\Macquarie}{School of Mathematical and Physical Sciences, Macquarie University, Balaclava Road, North Ryde, NSW 2109, Australia}
\newcommand{\CUBoulder}{Department of Physics, 390 UCB, University of Colorado, Boulder, CO 80309, USA}
\newcommand{\JPL}{Jet Propulsion Laboratory, California Institute of Technology, 4800 Oak Grove Drive, Pasadena, CA 91109, USA}
\newcommand{\MITEAPS}{Department of Earth, Atmospheric, and Planetary Sciences, Massachusetts Institute of Technology, Cambridge, MA 02139, USA}
\newcommand{\MITKavli}{Kavli Institute for Astrophysics and Space Research, Massachusetts Institute of Technology, Cambridge, MA 02139, USA}
\newcommand{\UCI}{Department of Physics \& Astronomy, The University of California, Irvine, Irvine, CA 92697, USA}
\newcommand{\Carnegie}{Earth and Planets Laboratory, Carnegie Institution for Science, 5241 Broad Branch Road, NW, Washington, DC 20015, USA}
\newcommand{\PSUICS}{Institute for Computational and Data Sciences, The Pennsylvania State University, University Park, PA 16802, USA}
\newcommand{\PSUCASt}{Center for Astrostatistics, 525 Davey Laboratory, The Pennsylvania State University, University Park, PA 16802, USA}
\newcommand{\Princeton}{Department of Astrophysical Sciences, Princeton University, 4 Ivy Lane, Princeton, NJ 08540, USA}
\newcommand{\IAS}{Institute for Advance Study, 1 Einstein Drive, Princeton, NJ 08540, USA}
\newcommand{\Tsinghua}{Department of Astronomy, Tsinghua University, Beijing 100084, China}
\newcommand{\FlatironCCA}{Center for Computational Astrophysics, Flatiron Institute, 162 Fifth Avenue, New York, NY 10010, USA}
\newcommand{\ETH}{ETH Zurich, Institute for Particle Physics \& Astrophysics, Zurich, Switzerland}
\newcommand{\UCO}{UC Observatories, University of California, Santa Cruz, CA 95064, USA}
\newcommand{\SantaCruz}{University of California, Santa Cruz}
\newcommand{\WMKO}{W.\ M.\ Keck Observatory, 65-1120 Mamalahoa Hwy, Kamuela, HI 96743, USA}
\newcommand{\SSL}{Space Sciences Laboratory, University of California, Berkeley, CA 94720, USA}
\newcommand{\UH}{Institute for Astronomy, University of Hawai‘i, 2680 Woodlawn Drive, Honolulu, HI 96822, USA}
\newcommand{\UCB}{Department of Astronomy, 501 Campbell Hall, University of California, Berkeley, CA 94720, USA}
\newcommand{\UCLA}{Department of Physics \& Astronomy, University of California Los Angeles, Los Angeles, CA 90095, USA}
\newcommand{\nexsci}{NASA Exoplanet Science Institute/Caltech-IPAC, MC 100-22, 1200 E.\ California Blvd., Pasadena, CA
91125, USA}
\newcommand{\COO}{Caltech Optical Observatories, California Institute of Technology, Pasadena, CA 91125, USA}
\newcommand{\Sydney}{Sydney Institute for Astronomy (SIfA), School of Physics, University of Sydney, NSW 2006, Australia}
\newcommand{\Kansas}{Department of Physics and Astronomy, University of Kansas, Lawrence, KS, USA}
\newcommand{\Warwick}{Physics Department, University of Warwick, Coventry CV4 7AL, United Kingdom}
\newcommand{\Yale}{Department of Astronomy, Yale University, New Haven, CT 06520, USA}
\newcommand{\ICL}{Astrophysics Group, Department of Physics, Imperial College London, Prince Consort Rd, London SW7 2AZ, UK}
\newcommand{\Schmidt}{Astrophysics \& Space Institute, Schmidt Sciences, New York, NY 10011, USA}
\newcommand{\Amsterdam}{University of Amsterdam}
\newcommand{\ND}{Department of Physics and Astronomy, University of Notre Dame, Notre Dame, IN 46556, USA}
\newcommand{\Geneva}{Observatoire Astronomique de l'Université de Genève, Chemin Pegasi 51, 1290 Versoix, Switzerland}

\title{The Compositions of Rocky Planets in Close-in Orbits Tend to be Earth-Like}

\correspondingauthor{Casey L. Brinkman}
\email{clbrinkm@hawaii.edu}

\author[0000-0002-4480-310X]{Casey L. Brinkman}
\altaffiliation{NSF Graduate Research Fellow}
\affiliation{Institute for Astronomy, University of Hawai`i, 2680 Woodlawn Drive, Honolulu, HI 96822, USA}

\author[0000-0002-3725-3058]{Lauren M. Weiss}
\affiliation{Department of Physics and Astronomy, University of Notre Dame, Notre Dame, IN, 46556, USA}

\author[0000-0001-8832-4488]{Daniel Huber}
\affiliation{Institute for Astronomy, University of Hawai`i, 2680 Woodlawn Drive, Honolulu, HI 96822, USA}
\affiliation{Sydney Institute for Astronomy (SIfA), School of Physics, University of Sydney, NSW 2006, Australia}

\author[0000-0001-7058-4134]{Rena A. Lee}
\altaffiliation{NSF Graduate Research Fellow}
\affiliation{Institute for Astronomy, University of Hawai`i, 2680 Woodlawn Drive, Honolulu, HI 96822, USA}

\author{Jared Kolecki}
\affiliation{Department of Physics and Astronomy, University of Notre Dame, Notre Dame, IN, 46556, USA}

\author[0009-0007-4721-9068]{Gwyneth Tenn}
\affiliation{Punahou School, 1601 Punahou St, Honolulu, HI 96822 USA}

\author[0000-0002-2696-2406]{Jingwen Zhang}
\altaffiliation{NASA FINESST Fellow}
\affiliation{Institute for Astronomy, University of Hawai`i, 2680 Woodlawn Drive, Honolulu, HI 96822, USA}

\author[0000-0002-0244-6650]{Suchitra Narayanan}
\altaffiliation{NSF Graduate Research Fellow}
\affiliation{Institute for Astronomy, University of Hawai`i, 2680 Woodlawn Drive, Honolulu, HI 96822, USA}

\author[0000-0001-7047-8681]{Alex S. Polanski}
\affiliation{Department of Physics and Astronomy, University of Kansas, Lawrence, KS, USA}

\author[0000-0002-8958-0683]{Fei Dai}
\affiliation{Institute for Astronomy, University of Hawai`i, 2680 Woodlawn Drive, Honolulu, HI 96822, USA}



\author[0000-0003-4733-6532]{Jacob L. Bean}
\affiliation{Department of Astronomy $\&$ Astrophysics, University of Chicago, 5640 S. Ellis Avenue, Chicago, IL 60637, USA}

\author[0000-0001-7708-2364]{Corey Beard}
\affiliation{Department of Physics $\&$ Astronomy, University of California Irvine, Irvine, CA 92697, USA}
\altaffiliation{NASA FINESST Fellow}

\author[0000-0003-2404-2427]{Madison Brady}
\affiliation{Department of Astronomy $\&$ Astrophysics, University of Chicago, 5640 S. Ellis Avenue, Chicago, IL 60637, USA}

\author[0009-0008-9808-0411]{Max Brodheim}
\affiliation{\WMKO}

\author{Matt Brown}
\affiliation{\WMKO}


\author[0000-0003-1125-2564]{Ashley Chontos}
\affiliation{Department of Astrophysical Sciences, Princeton University, 4 Ivy Lane, Princeton, NJ 08544, USA}

\author[0009-0000-3624-1330]{William Deich}
\affil{\UCO}

\author[0009-0002-2419-8819]{Jerry Edelstein}
\affiliation{\SSL}

\author[0000-0003-3504-5316]{Benjamin J.\ Fulton}
\affiliation{Department of Astronomy, California Institute of Technology, Pasadena, CA 91125, USA}


\author[0000-0002-8965-3969]{Steven Giacalone}
\affiliation{Department of Astronomy, California Institute of Technology, Pasadena, CA 91125, USA}

\author[0009-0004-4454-6053]{Steven R.\ Gibson}
\affiliation{\COO}

\author[0000-0003-0742-1660]{Gregory J. Gilbert}
\affiliation{Department of Physics \& Astronomy, University of California Los Angeles, Los Angeles, CA 90095, USA}




\author[0000-0003-1312-9391]{Samuel Halverson}
\affil{\JPL}

\author[0000-0002-9305-5101]{Luke Handley}
\affiliation{Department of Physics \& Astronomy, University of California Los Angeles, Los Angeles, CA 90095, USA}
\affiliation{Department of Astronomy, California Institute of Technology, Pasadena, CA 91125, USA}

\author[0000-0002-7648-9119]{Grant M.\ Hill}
\affiliation{\WMKO}

\author[0000-0002-5034-9476]{Rae Holcomb}
\affiliation{Department of Physics \& Astronomy, University of California Irvine, Irvine, CA 92697, USA}

\author[0000-0002-6153-3076]{Bradford Holden}
\affil{\UCO}

\author[0000-0002-5812-3236]{Aaron Householder}
\affiliation{\MITEAPS}
\affiliation{\MITKavli}

\author[0000-0001-8638-0320]{Andrew W. Howard}
\affiliation{Department of Astronomy, California Institute of Technology, Pasadena, CA 91125, USA}

\author[0000-0002-0531-1073]{Howard Isaacson}
\affiliation{Department of Astronomy, 501 Campbell Hall, University of California, Berkeley, CA 94720, USA}





\author{Stephen Kaye}
\affiliation{\COO}

\author[0000-0003-2451-5482]{Russ R. Laher}
\affiliation{NASA Exoplanet Science Institute/Caltech-IPAC, MC 100-22, 1200 E California Blvd, Pasadena, CA 91125, USA}

\author[0009-0004-0592-1850]{Kyle Lanclos}
\affiliation{\WMKO}





\author[0000-0001-7664-648X]{J. M. Joel Ong}
\altaffiliation{NASA Hubble Fellow}
\affiliation{Institute for Astronomy, University of Hawai`i, 2680 Woodlawn Drive, Honolulu, HI 96822, USA}

\author[0009-0008-4293-0341]{Joel Payne}
\affiliation{\WMKO}

\author[0000-0003-0967-2893]{Erik A. Petigura}
\affiliation{Department of Physics \& Astronomy, University of California Los Angeles, Los Angeles, CA 90095, USA}

\author[0000-0001-9771-7953]{Daria Pidhorodetska}
\affil{Department of Earth and Planetary Sciences, University of California, Riverside, CA 92521, USA}

\author[0000-0003-0512-5489]{Claire Poppett}
\affiliation{\SSL}




\author[0000-0001-8127-5775]{Arpita Roy}
\affiliation{\Schmidt}

\author[0000-0003-3856-3143]{Ryan Rubenzahl}
\affiliation{Department of Astronomy, California Institute of Technology, Pasadena, CA 91125, USA}


\author[0000-0003-2657-3889]{Nicholas Saunders}
\altaffiliation{NSF Graduate Research Fellow}
\affiliation{Institute for Astronomy, University of Hawai`i, 2680 Woodlawn Drive, Honolulu, HI 96822, USA}

\author[0000-0002-4046-987X]{Christian Schwab}
\affil{\Macquarie}

\author[0000-0003-4526-3747]{Andreas Seifahrt}
\affiliation{Department of Astronomy $\&$ Astrophysics, University of Chicago, 5640 S. Ellis Avenue, Chicago, IL 60637, USA}

\author[0000-0003-3133-6837]{Abby P.\ Shaum}
\affiliation{Department of Astronomy, California Institute of Technology, Pasadena, CA 91125, USA}
\author[0009-0007-8555-8060]{Martin M.\ Sirk}
\affiliation{\SSL}

\author{Chris Smith}
\affiliation{\SSL}

\author[0000-0001-7062-9726]{Roger Smith}
\affiliation{\COO}

\author[0000-0001-7409-5688]{Guðmundur Stefánsson} 
\affiliation{Anton Pannekoek Institute for Astronomy, University of Amsterdam, Science Park 904, 1098 XH Amsterdam, The Netherlands}

\author[0000-0002-4410-4712]{Julian St\"urmer}
\affiliation{Landessternwarte, Zentrum f\"ur Astronomie der Universit\"at Heidelberg, K\"onigstuhl 12, D-69117 Heidelberg, Germany}

\author{Jim Thorne}
\affiliation{\WMKO}

\author[0000-0002-1845-2617]{Emma V. Turtelboom}
\affiliation{Department of Astronomy, 501 Campbell Hall, University of California, Berkeley, CA 94720, USA}

\author[0000-0003-0298-4667]{Dakotah Tyler}
\affiliation{Department of Physics and Astronomy, University of California, Los Angeles, CA 90095, USA}

\author{John Valliant}
\affiliation{\WMKO}


\author[0000-0002-4290-6826]{Judah Van Zandt}
\affiliation{Department of Physics \& Astronomy, University of California Los Angeles, Los Angeles, CA 90095, USA}

\author[0000-0002-6092-8295]{Josh Walawender}
\affiliation{\WMKO}




\author[0000-0001-7961-3907]{Samuel W.\ Yee}
\affiliation{Center for Astrophysics \textbar \ Harvard \& Smithsonian, 60 Garden Street, Cambridge, MA 02138, USA}
\altaffiliation{Heising-Simons Foundation 51 Pegasi b Postdoctoral Fellow}

\author{Sherry Yeh}
\affiliation{\WMKO}

\author{Jon Zink}
\affiliation{Department of Astronomy, California Institute of Technology, Pasadena, CA 91125, USA}

\begin{abstract}

Hundreds of exoplanets between 1-1.8 times the size of the Earth have been discovered on close in orbits. However, these planets show such a diversity in densities that some appear to be made entirely of iron, while others appear to host gaseous envelopes. To test this diversity in composition, we update the masses of 5 rocky exoplanets (HD 93963 A b, Kepler-10 b, Kepler-100 b, Kepler-407 b, and TOI-1444 b) and present the confirmation of a new planet (TOI-1011) using 187 high precision RVs from Gemini/MAROON-X and Keck/KPF. Our updated planet masses suggest compositions closer to that of the Earth than previous literature values for all planets in our sample. In particular, we report that two previously identified ``super-Mercuries'' (Kepler-100 b and HD 93963 A b) have lower masses that suggest less iron-rich compositions. We then compare the ratio of iron to rock-building species to the abundance ratios of those elements in their host stars. These updated planet compositions do not suggest a steep relationship between planet and host star compositions, contradictory to previous results, and suggest that planets and host stars have similar abundance ratios.

\end{abstract}


\section{Introduction}
\label{sec:intro}

In our effort to contextualize the properties of the Earth amongst the thousands of extra-solar planets discovered, one of the most challenging properties to compare is planet composition. Rocky planet compositions affect habitability \citep{2014ebi..conf..2.1M, 2020plas.book..449K, 2018haex.bookE..76D}, and can inform planet formation scenarios (\citealt{scora2020, Adibekyan2021, 2024arXiv240908361B}). To characterize a planet's composition and determine if it is Earth-like, iron-rich like Mercury, or hosts a thick atmosphere unlike any small planets in our solar system, we need precisely measured masses and radii. Radial velocity (RV) measurements of transiting planets are one of the best tools to better understand the masses, densities, and compositions of other worlds. 

Our ability to determine composition of small planets from mass and radius measurements is based on an assumption of a purely ``rocky'' composition. If a planet hosts a water layer or volatile envelope, it becomes impossible to constrain the relative fraction of iron core to rocky mantle for a planet, even with a precisely measured bulk density \citep{Valencia2007, RogersSeager2010}. However, for planets that are sufficiently small ( R$<$1.5 R$_{\oplus}$) that orbit very close to their host star (P < 30 days) and receive large stellar flux, we can assume they are unlikely to have significant atmospheres \citep{2017ApJ...847...29O} or water/ice layers \citep{2017MNRAS.472..245L} due to photo-evaporation. This leaves silicate rock and  iron as the two primary components of short-period super-Earths. To first order, we can express the composition of rocky planets using the fraction of the planet's mass that is iron, or its Core Mass Fraction (CMF). 

The masses and radii of small exoplanets suggest a transition between primarily rocky (super-Earth) and gas-enveloped (sub-Neptune) planets at approximately 1.5 $R_{\oplus}$ \citep{2014ApJ...783L...6W, 2015ApJ...801...41R, 2017AJ....154..109F}, with planets smaller than 1.5 $R_{\oplus}$ often having compositions consistent with Earth-like iron-to-silicate ratios \citep{2015ApJ...800..135D}. The mass and radius measurements for super-Earths, however, indicate a wide diversity of densities amongst these planets---far more diverse than we observe for small planets in our own solar system \citep{2014ApJS..210...20M, 2016ApJ...822...86M, 2019ApJ...883...79D}. These densities suggest the interior compositions of Earth and super-Earth sized planets could potentially vary from entirely made of silicate rock, to predominantly made of iron \citep{2019NatAs...3..416B}, with high-molecular-mass atmospheres possible \citep{2017AJ....154..232A, 2021ApJ...909L..22K}. 

To better understand the compositions of rocky worlds we can place these planets in the context of their host star. Planets are born from the same primordial nebular material as their host star, and it is intuitive to assume that the relative chemical abundances of iron and rock-building elements between star and planet would be similar. While some studies have explicitly assumed similar elemental abundance ratios for stars and planets \citep{Dorn2015} others have tried to test it (\citealt{Plotnykov2020, Adibekyan2021, Schulze2021, 2024arXiv240908361B}). Most studies found that the uncertainties---especially in mass---are too large in most cases to draw definitive conclusions about the compositions of individual rocky planet and host star systems \citep{Plotnykov2020, Schulze2021}. Additionally, many of the best-characterized rocky planets (such as the TRAPPIST-1 system) orbit stars too cool for individual abundance measurements of Fe and Mg. 

This apparent diversity of compositions is based on a small sample of rocky planets, with most having large uncertainties on their mass measurements. Rocky planets have small radii and low masses, which are inherently more difficult to measure, and only a handful have masses and radii measured to within 10$\%$ \citep{2017NatAs...1E..56G, 2019ApJ...883...79D, 2020MNRAS.491.2982E, 2021PSJ.....2....1A, 2021Sci...371.1038T, 2021A&A...649A.144S, 2021NatAs...5..775D, Brinkman2023B, 2023A&A...677A..33B}. Many of the planets that appear to have compositions most dissimilar to the Earth, such as high-density ``super-Mercuries,'' have large uncertainties in their mass measurements. Updating these masses with high-precision and high-cadence RV measurements will help to characterize the compositions of these worlds.

To address these issues and better understand the compositional diversity of rocky planets, we report 187 high-precision RVs for 6 rocky planets, including confirmation of a previously unconfirmed planet TOI-1011 b. We homogeneously update the stellar masses and radii using isochrones, and report updated masses and radii for each planet. We then compute the CMF and where possible analyze the composition of each planet in relation to that of its host star.  

\section{Sample Selection}

To select targets for our RV survey we use both objective and subjective criteria to sample a wide variety of apparent compositions. First, we selected planets between 1-1.8 R$_{\oplus}$ that have published RVs using the NASA Exoplanet Archive (queried 4/18/2024, \cite{2013PASP..125..989A}), with either a published value for the mass or an upper limit. 

We then selected only short period (P$<$10 days) and Ultra-short period (P$<$ 1 day) planets, because MAROON-X and KPF are both recently commissioned instruments whose long-term stability is still being characterized. As such, our data acquision strategy was to collect all of our RVs on a particular star within one run on MAROON-X (order of 1 week) and as many as possible per night with KPF. To mitigate long-term stability issues, we use a different vertical offset for each run of MAROON-X data and each night of KPF data (colloquially called the ``Floating Chunk'' method).

Another large consideration in selecting the sample of planets was available telescope resources. We wanted to produce a sample of planets with masses measured with a fractional precision of 10$\%$, so we selected planets that could meet this threshold with the number of RVs we are able to collect over three years of observing. To determine this, we simulated RV measurements with fractional uncertainties of 1 m/s to mimic those collected by MAROON-X and KPF, and added those to the existing RV datasets for a particular star and recovered the best-fit semi-amplitude and 1$\sigma$ uncertainty. We then determined how many RVs would be necessary to measure a semi-amplitude for the planet to 10$\%$ uncertainty (using the previously published semi-amplitude) and then calculated the necessary exposure time to achieve 1 m/s precision on that star with either MAROON-X or KPF.  


We selected 9 planets for our survey that meet these criteria and represent planets that range from low-density planets likely to host gaseous envelopes, to Earth-like, to high-density super-Mercuries: Kepler-10 b, Kepler-100 b, Kepler-407 b, Kepler-93 b, Kepler-99 b, TOI-561 b, TOI-1444, HD 93963 A b, and GJ 3929 b. The RV analysis for TOI-561 b was presented separately in \citep{Brinkman2023B}. For three of these planets (Kepler-93 b, Kepler-99 b, and GJ 3929 b) the KPF RVs are still preliminary and will be published at a later date when the wavelength solution is stable enough that a night-to-night offset is not needed. 

We include one additional unconfirmed planet in our sample: TOI-1011. This planet was discovered as part of a search for RV signals in archival HARPS-N RVs around stars flagged with planet candidates from TESS. Because the radius, preliminary mass measurement, and orbital period from this was consistent with our survey criteria we added TOI-1011 to our sample.

\section{Observations}
\subsection{HIRES}
Our analysis incorporates RVs from the High Resolution Echelle Spectrograph (HIRES) on the W. M. Keck Observatory 10m telescope Keck-I on Maunakea, Hawai`i \citep{1994SPIE.2198..362V}. We observed TOI-561, Kepler-10, Kepler-100, and Kepler-407 with HIRES from February 2021-January 2023\footnote{Telescope time was allocated by University of Hawaii}. HIRES is a well-characterized spectrograph, with demonstrated stability \citep{2010ApJ...721.1467H}. 

We used the standard California Planet Search (CPS) data reduction pipeline as described in \cite{2010ApJ...721.1467H}. This method uses an iodine cell mounted in front of the slit in order to provide a provide a wavelength reference \citep{1992PASP..104..270M}.  Sky subtraction was performed as part of the raw reduction through the use of a $14\arcsec.0$ long slit in order to spatially resolve the sky with respect to the seeing-limited point-spread function (full-width half-max $\approx$ $1\arcsec.0$). Measuring the RVs requires characterizing the PSF of the spectrometer, which is time-variable due primarily to changing seeing and weather.  The CPS Doppler routine involves forward-modeling the iodine-imprinted spectrum of a star as the combination of a library iodine spectrum and a velocity-shifted, iodine-free, PSF-deconvolved template spectrum of the target star, the combination of which is then convolved with the best-fit PSF. To deconvolve the PSF from the iodine-free template, we observed rapidly-rotating B stars with the iodine cell in the light path immediately before and after the template, effectively sampling the PSF at the time of the template in the iodine absorption profiles.

\subsection{MAROON-X}
MAROON-X is a new fiber-fed spectrograph mounted on the 8.1 meter Gemini-North telescope on Maunakea, Hawai`i. It operates in the red-optical (500-920nm) with resolving power R$\approx$85,000, and uses both red and blue arms to get two radial velocity measurements per exposure \citep{2016SPIE.9908E..18S, 2018SPIE10702E..6DS, 2020SPIE11447E..1FS}. MAROON-X has demonstrated an intra-night stability of 30 cm/s, and has been used to measure some of the most precise masses for rocky planets in the literature to date \citep{2021Sci...371.1038T, 2022AJ....163..168W, Brinkman2023B}. We observed TOI-561, TOI-1011, Kepler-10, Kepler-100, and Kepler-407 with MAROON-X between January 2021 and December 2023. 

Our observations used the simultaneous calibration fiber of MAROON-X, which allows for a robust order-by-order drift correction to sub-m/s precision. The raw data was reduced using a custom pipeline based on CRIRES \citep{2010Msngr.140...41B}, and RVs were computed using \texttt{SERVAL} \citep{2018A&A...609A..12Z}. A full description of MAROON-X data reduction can be found in \cite{2022AJ....163..168W}.

Exposure times were chosen to achieve a precision of 1 m/s on our RVs for most targets (and 1.5 m/s on Kepler-407 due to large integration times). We estimated the necessary exposure time by scaling the demonstrated RV precision from 51 Pegasi (21 cm/s photon-limited precision for a 120s exposure) using $\frac{\sigma_{RV}}{\sigma_{RV0}}=\frac{SNR_{0}}{SNR}=\sqrt{\frac{F_{0}\times t_{0}}{F \times t}}$. Most of our targets were similar spectral types to 51 Pegasi (G-type), and we found strong agreement between our estimated and actual RV precision. 

\subsection{KPF}
The Keck Planet Finder \citep{10.1117/12.3017841} is a fiber-fed spectrograph that observes in the optical regime with a wavelength coverage of 445--870 nm and a resolving power of R$\approx$95,000. This wavelength range is broken into a green channel (445--600 nm) and a red channel (600--870 nm). Additionally, a UV spectrometer (385--405 nm) monitors CaII H $\&$ K, the emission cores of which are commonly used as indicators for magnetic activity in the stellar chromosphere. With a goal Doppler precision of 30 cm/s, KPF is exceptionally well suited to measure the small semi-amplitudes of rocky planets \citep{2024AJ....168..101D}. 

We obtained spectra of Kepler-10, Kepler-100, TOI-1444, and HD 93963 with KPF across 2023 A and 2023 B (May 2023-January 2024). We used the publicly available exposure time calculator \footnote{\url{https://github.com/California-Planet-Search/KPF-etc}} to estimate the exposure time necessary to produce RVs with 1 m/s precision. The spectra were reduced with the KPF Data Reduction Pipeline (DRP), also available on Github\footnote{\url{https://github.com/Keck-DataReductionPipelines/KPF-Pipeline}}. 

The wavelength calibration for KPF includes Th-Ar and U-Ne lamps, a Laser Frequency Comb, a Fabry-P\'erot Etalon, and the Solar Calibrator \citep{KPFSoCal}. Wavelength calibration of KPF DRP is still in development at the time of writing, and nightly offsets in the wavelengths are anticipated. To mitigate this effect, we give each night of KPF data a different vertical offset ($\gamma$) in the RV model. As such, only KPF nights with high cadence observations for each star are useful, although we anticipate that future work on the wavelength solution will allow comparison of RVs across multiple nights.

\subsection{HARPS-North}
Our analysis of Kepler-10 and TOI-1011 also incorporate published RVs from the HARPS-N spectrograph installed on the 3.6 meter Telescopio Nazionale Galileo (TNG) at the Observatorio Roque de Los Muchachos in La Palma, Spain. These RVs are taken from \cite{bonomo2023}. 

\section{Methodology}
\subsection{RV Fitting and Planet Masses}

We used the open source python package \texttt{RadVel} \citep{2018PASP..130d4504F} to model the RVs. We measured the mass of each planet by modeling the RVs for a Keplerian orbit, in which the RV curve is described by the orbital period (P), inferior conjunction time (T$_{c}$), eccentricity (e), argument of periastron, and RV semi-amplitude (K) of each planet. In all instances we used the orbital period and conjunction time---along with their uncertainties---from the photometric transit fit as Gaussian priors on these parameters. For most planets we assumed circular orbits  \citep{2013ApJ...774L..15D, 2019AJ....157...61V, 2019AJ....157..198M, 2021AJ....162...55Y}, and checked that this was consistent with previous literature fits. We then allowed eccentricity to vary and found in all instances that the resulting RV fit was consistent with a fixed zero eccentricity. 

We included two additional terms per dataset to fit the RVs: a zeropoint offset ($\gamma$) and an RV jitter term ($\sigma_{j}$). Jitter accounts for additional Gaussian noise that can be astrophysical in origin, or can come from systematics of the spectrograph. This additional uncertainty was added in quadrature with the intrinsic uncertainties on the RVs during our optimization of the likelihood function and Markov Chain Monte Carlo (MCMC) analysis. The likelihood function used in \texttt{RadVel} is:
\begin{equation}
    \ln(\mathcal{L}) =-\sum\limits_{i} \frac{(v_{i}-v_{m}(t_{i}))^{2}}{2(\sigma_{i}^{2}+\sigma_{\rm{jit}}^{2})} - \ln \sqrt{2\pi (\sigma_{i}^{2}+\sigma_{\rm{jit}}^{2})} 
\end{equation}
where $\mathcal{L}$ is the Likelihood, $v_{i}$ and $\sigma_{i}$ are the $i$th radial velocity measurement and its associated uncertainty, $v_{m}(t_{i})$ is the Keplerian model radial velocity at time $t_{i}$, and $\sigma_{\rm{jit}}$ is the jitter estimate. After optimizing for the maximum likelihood fit, we ran \texttt{RadVel}'s built-in MCMC algorithm \citep{2013PASP..125..306F} to estimate the uncertainty in the model parameters, and to explore the covariance between parameters.

Once we found the best-fit RV model for each system, we calculated the planet mass (m) using the best-fit semi-amplitude (K), orbital period (P), inclination (i), eccentricity (e), and stellar mass (M) using the following relation \citep{2010exop.book...27L}: 

\begin{equation}
    K=\frac{28.4329}{\sqrt{1-e^{2}}}\frac{m \sin(i)}{M_{J}}\left(\frac{m + M}{M_{\odot}}\right)^{-2/3}\left(\frac{P}{ \text{1 year}}\right)^{-1/3}
\end{equation}

We used our updated stellar masses (Table \ref{tab:stars}, along with literature values for orbital period and inclination for all previously confirmed systems (Table insert). We used orbital period and inclination from our photometric analysis for TOI-1011 b. 

\subsection{Planet Compositions}

Using the mass and radius of each plane we calculated the CMF, a measure of the mass fraction of the iron core to the total planet mass \footnote{For reference, the Earth is $32.5\%$ iron by mass, giving it a CMF of 0.325.} \citep[e.g.][]{2007ApJ...669.1279S, 2014ApJ...787..173H, 2016ApJ...819..127Z}. This is a measure of the minimum CMF for the planet, and is only an accurate CMF in the absence of a volatile envelope or water layer. Because we have selected only short and ultra-short period planets with R$<$1.5 R$_{\oplus}$, they are less likely to host these low density components \citep{2017ApJ...847...29O, 2017MNRAS.472..245L}. However, this is not universally true, and there are ultra-hot super-Earths likely to host gaseous envelopes, such as TOI-561 b \citep{2021MNRAS.501.4148L, Brinkman2023B}. 

We used \texttt{SuperEarth} \citep{2006Icar..181..545V, Plotnykov2020} to model the interior composition of each planet. The package solves equations of state for iron and various rock-building minerals to match the mass and radius values provided. To first order, the code assumes the planet has two primary, differentiated layers (an iron core and a rocky mantle). \texttt{SuperEarth} then refines this approximation by assuming a four-layer mantle composition (upper mantle, transition zone, lower mantle, and lower-most mantle) like that of the Earth, distinguished by the mineral phase boundaries determined by the pressure and temperature of the mantle. We also assume an iron fraction of 0.1 in the mantle, and used the default silicate inclusion in the core of 0.0 \citep{Plotnykov2020}. We used our newly measured masses along with literature values for radius to compute the CMF for each planet in our sample. 

To compute CMF uncertainties, we drew 1000 values for the mass and radius of each planet from Gaussian distributions centered on best-fit values with 1$\sigma$ error bars. We then propagated these values through \texttt{SuperEarth} and took the standard deviation of the resulting CMF distribution, (after ensuring that distribution was also Gaussian). We assumed no correlation between planet mass and radius, although there likely is one, and therefore our Monte Carlo draws represent conservative uncertainty estimates.

\subsection{Host Star Properties}
\subsection{Atmospheric Parameters}

For most stars in our sample we use literature values for T$_{\mathrm{eff}}$, [M/H], [$\alpha$/Fe], and log($g$), but for TOI-1011 we measure our own using a single HARPS-N spectra obtained on March 20 2019 (T$_{exp}$=1165 s, SNR=92 at $\lambda$=530 nm). This was done using the code \texttt{moogplusplus}\footnote{\url{https://github.com/kolecki4/moogplusplus}} (Kolecki et al. in prep). $T_{\textrm{eff}}$ and $\log{(g)}$ were calculated by fitting photometry to a grid of MIST isochrones. These parameters are used to interpolate an appropriate model atmosphere from a pre-calculated grid of PHOENIX atmospheres \citep{Husser+2013}. 

\texttt{moogplusplus} then fits abundances using line-by-line spectral synthesis, performing a $\chi^2_\nu$ minimization fit to the observed data for spectral lines of a given element. Perturbing the abundance until the residuals are sufficiently minimized results in a best-fit synthetic spectrum for each observed line feature of a given element. Each individual synthetic spectral line fit has its own unique abundance value, resulting in a distribution of abundance values. 

By taking the median of this distribution, we can report a single value for the abundance of the chosen element in the star. With the HARPS-N spectrum of TOI-1011, we were able to successfully fit 388 lines of Fe, resulting in median values of A(Fe)=7.53 $\pm$ 0.03 and [Fe/H]=0.03 $\pm$ 0.01. This uncertainty on [Fe/H] represents the intrinsic uncertainty, and we inflate it to 0.04 dex to account for potential systematic uncertainties, which is the median uncertainty of other [Fe/H] values in our sample and in agreement with \cite{2012ApJ...757..161T}. \texttt{moogplusplus} calculates [$\alpha$/Fe] as the average of the abundances of Ca and Ti, resulting a value for TOI-1011 of [$\alpha$/Fe] = 0.05 $\pm$ 0.04.

\subsubsection{Masses and Radii}
\label{sec:starmass}
\begin{figure*}
    \centering
    \includegraphics[width=1.0\textwidth]{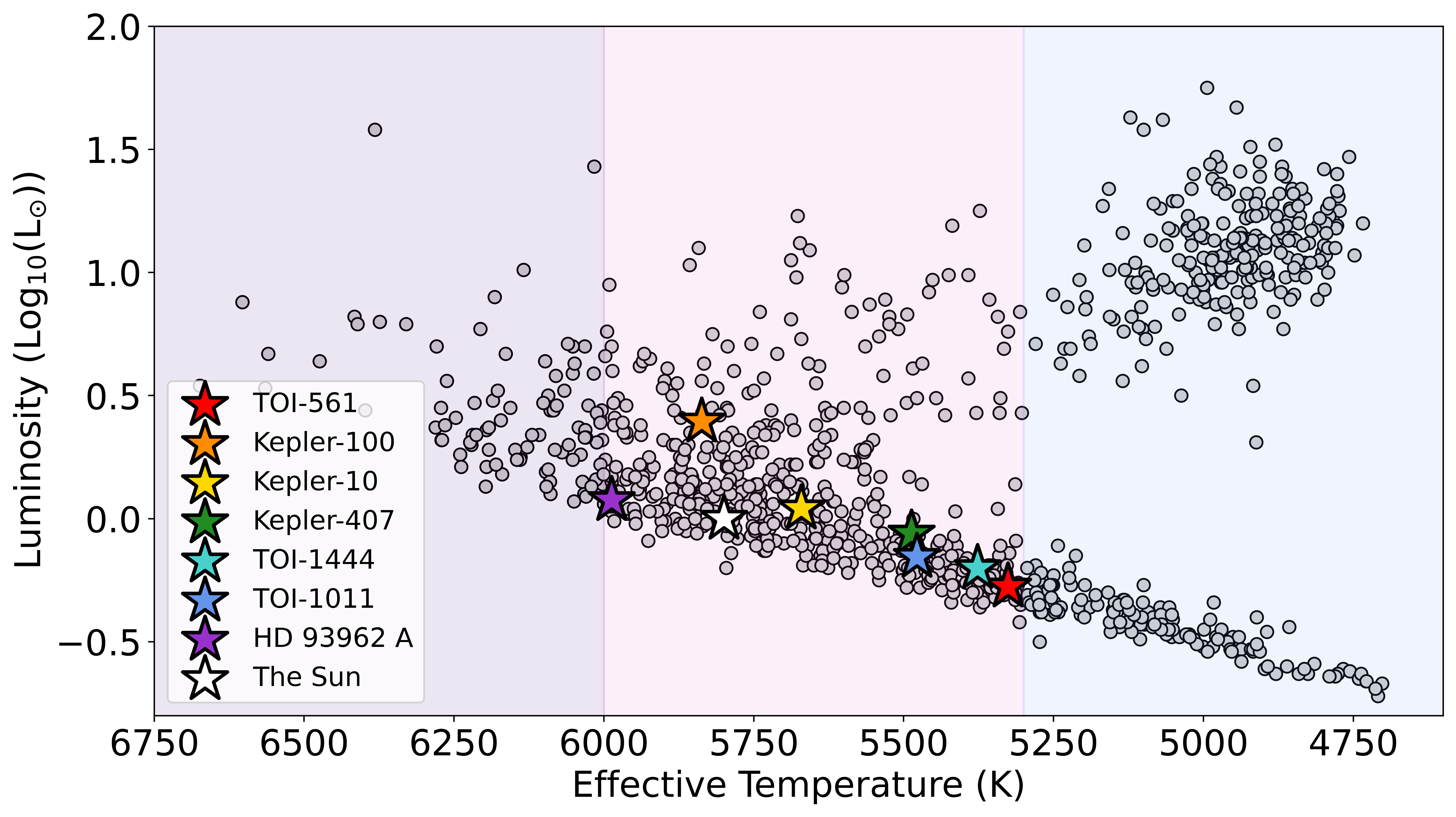}
    \caption{Effective temperature vs luminosity is shown for all stars in our sample (colored star symbols) along with a larger sample of exoplanet host stars from the SPOCS catalogue \citep{2018ApJS..237...38B}. Stars of spectral type G are shown in the pink shaded region (center), those of type F in the purple shaded region (left), and K in the blue shaded region (right).}
    \label{fig:hr}
\end{figure*}

To create a more homogeneously characterized sample of planets, we updated the masses and radii of each host star in our sample using \texttt{isoclassify} \citep{2017ApJ...844..102H}. We used the ``direct mode'' to calculate the luminosity of each star using T$_{\mathrm{eff}}$, [M/H], and log($g$), along with the Gaia DR3 parallax, 2MASS K-band magnitude, a 3D dust map and bolometric corrections. We then used these derived luminosities, T$_{\mathrm{eff}}$, and [M/H] in the ``grid mode'' of \texttt{isoclassify} to infer the mass and radius of each star from a grid of MIST isochrones \citep{2016ApJ...823..102C}. Our newly derived luminosities, masses and radii of the planet-hosting stars are listed in Table \ref{tab:stars} and shown in Figure \ref{fig:hr}

We use literature spectroscopic values for T$_{\mathrm{eff}}$, [M/H], and log($g$) from Polanski et al. in prep (Kepler-10, Kepler-100, TOI-561, and TOI-1444), \citealt[(Kepler-407)]{2018ApJS..237...38B}, and \citealt[]{2022A&A...667A...1S} (HD 93963 A). 


\subsubsection{Compositions}
\label{sec:starcmf}

To compare the composition of our planets with that of their host star, we must express the stellar abundances and planet compositions in equivalent quantities. For stars with measured iron and magnesium abundances, we computed the mass ratio of Fe/Mg from these abundance measurements and then computed the stellar equivalent value of planet CMF. Stellar abundance measurements are given in the form:
\begin{equation}
  [E/H]=\log_{10}(\frac{n(E)/n(H)_{*}}{n(E)/n(H)_{\odot}})
  \label{eq:abundnce}
\end{equation}
where n(E)/n(H)$_{*}$ is the number density of an element E relative to hydrogen. To turn this into an absolute number density for the star (not relative to the Sun), we used the number density for each element in the Sun relative to hydrogen, given as:
\begin{equation}
    A(E)=12 + \log_{10}(n(E)/n(H)_{\odot}).
\end{equation}
Using A(Fe)=7.46 $\pm$ 0.04, A(Mg)=7.55 $\pm$ 0.03, and A(Si)=7.51 $\pm$ 0.03 \citep{2021A&A...653A.141A}, we solved for the number density of these three elements relative to hydrogen (n(E)/n(H)$_{\odot}$). We then used these values to solve for n(E)/n(H)$_{*}$ in Equation \ref{eq:abundnce}. With values for the number density of each element, we then calculated the mass of each of these elements relative to hydrogen using the atomic weights of each species (55.85 u for Fe and 24.3 u for Mg). This allows us to calculate the mass ratio Fe/Mg for each star.

We then translated this Fe/Mg ratio in the host star into an equivalent value for ``Core Mass Fraction'' (CMF$_{*}$), where the core here is not the core of the star but an expression of iron mass to total iron and rock-building element mass (assuming the same ratio of Mg, Si, and O as the planet mineral composition). This is done through a function in \texttt{SuperEarth} that uses the inverse process of computing Fe/Mg ratio from CMF for the planets as described above. This allows us to compare equivalent quantities for star and planet composition. 

We calculated the uncertainties on CMF$_{*}$ by drawing 1000 values of [E/H] from Gaussian distributions centered on our measurements with 1$\sigma$ widths, and then computing the Fe/Mg mass ratio and CMF$_{*}$ for each draw and taking the standard deviation of the resulting CMF$_{*}$ distributions for each star.

\section{Individual Systems}
\begin{table*}
\begin{center}
\begin{tabular}{|c|c|c|c|c|c|c|c|c|c|}
\hline
Star Name  & Radius           & Mass             & Luminosity       & Teff                & log(g)        & [Fe/H]           & [Mg/H]          & [$\alpha$/Fe]    & CMF$_{*}$ \\ 
           & R$_{\odot}$      & M$_{\odot}$      & L$_{\odot}$      & K                   &               & dex              &  dex            & dex              &  \\ \hline
HD 93963 A & 1.03  $\pm$ 0.01 & 1.09 $\pm$ 0.02  & 1.19 $\pm$ 0.05  & 5987 $\pm$ 64 $^{S}$& 4.5 $\pm$ 0.1 & 0.1 $\pm$ 0.04   & 0.08 $\pm$ 0.06 & -                & 0.31\\\hline
Kepler-10  & 1.06  $\pm$ 0.06 & 0.89 $\pm$ 0.05  & 1.1 $\pm$ 0.1    & 5671 $\pm$ 100 $^{P}$& 4.4 $\pm$ 0.1 & -0.17 $\pm$ 0.04 & -0.07           & 0.10 $\pm$ 0.04  & 0.25 \\\hline
Kepler-100 & 1.51 $\pm$ 0.05  & 1.13 $\pm$ 0.05  & 2.48 $\pm$ 0.07  & 5837 $\pm$ 100 $^{P}$& 4.1 $\pm$ 0.1 & 0.10 $\pm$ 0.04  & 0.05            & -0.02 $\pm$ 0.04 & 0.33 \\\hline
Kepler-407 & 1.04 $\pm$ 0.05  & 0.99 $\pm$ 0.05 & 0.78 $\pm$ 0.05   & 5487 $\pm$ 100 $^{B}$& 4.3 $\pm$ 0.1 & 0.35 $\pm$ 0.05  & 0.33 $\pm$ 0.07 & -0.03 $\pm$ 0.04 & 0.40 \\\hline
TOI-1011   & 0.92 $\pm$ 0.03  & 0.91 $\pm$ 0.03 & 0.70 $\pm$ 0.03   & 5475 $\pm$ 84        & 4.5 $\pm$ 0.05& 0.03 $\pm$ 0.04  &  -              & 0.05 $\pm$ 0.04  & - \\\hline
TOI-1444   & 0.91 $\pm$ 0.03  & 0.88 $\pm$ 0.05 & 0.63 $\pm$ 0.02  & 5377 $\pm$ 100 $^{P}$& 4.4 $\pm$ 0.1 & 0.04 $\pm$ 0.04  & 0.04 $\pm$ 0.05 & 0.01 $\pm$ 0.04  & 0.29  \\\hline
\end{tabular}

\caption{Uncertainties on CMF are all 0.03. All stellar masses and radii were measured homogeneously using the Teff, [Fe/H], and log(g) listed here. Teff, log(g), [Fe/H], and [Mg/H] for HD 93963 A are from \cite{2022A&A...667A...1S}, those for Kepler-407 are from \cite{2018ApJS..237...38B}, and those for TOI-1011 are measured here. Kepler-10, Kepler-100, and TOI-1444 have parameters measured from Polanski et al. (in prep) and also found in Brinkman et al. (Submitted). Quoted stellar parameters and precisions do not include potential systematic errors from different model grids \citep{2022ApJ...927...31T}.}
\label{tab:stars}

\end{center}

\end{table*}

\begin{table*}
\begin{center}
    
\begin{tabular}{|c|c|c|c|c|c|c|}
\hline
 Planet Name  &  Orbital Period                  & R$_{p}$/R$_{*}$          & Planet Radius          & Semi-Amplitude & Planet Mass & Core Mass Fraction \\ 
              &  Days                            &                          & R$_{\oplus}$           & m/s            & M$_{\oplus}$ &                    \\ \hline

 HD 93963 A b &  1.037611(9)  & 0.0131 $\pm$ 0.0006 $^{P}$& 1.49 $\pm$ 0.04 & 2.56 $\pm$ 0.99 & 4.31 $\pm$ 1.66 & 0.33 $\pm$ 0.42 \\ \hline
 Kepler-10 b  &  0.8374907(2)      & 0.01268$\pm$0.00004 $^{D}$& 1.47 $\pm$ 0.03 & 2.58 $\pm$ 0.24 & 3.58 $\pm$ 0.33 & 0.16 $\pm$ 0.20 \\ \hline
 Kepler-100 b &  6.88734(7)         & 0.008062$\pm$0.001  $^{M}$& 1.34 $\pm$ 0.12  & 1.25 $\pm$ 0.15 & 4.01 $\pm$ 0.47 & 0.59 $\pm$ 0.30 \\ \hline 
 Kepler-407 b &  0.6693124(6) & 0.010404$\pm$0.0002 $^{M}$ & 1.19 $\pm$ 0.05 & 1.41 $\pm$ 0.39 & 1.93 $\pm$ 0.50 & 0.35 $\pm$ 0.32 \\ \hline
 TOI-1011 b   &  2.470498(7)  & 0.0143 $\pm$ 0.0006 &1.45 $\pm$ 0.05 &  2.03 $\pm$ 0.30 & 4.04 $\pm$ 0.59 & 0.33 $\pm$ 0.19 \\ \hline
 TOI-1444 b   &  0.470269(4)      & 0.01427 $\pm$ 0.0004 $^{P}$&1.42 $\pm$ 0.04  & 2.85 $\pm$ 0.37 & 3.34 $\pm$ 0.43 & 0.22 $\pm$ 0.17 \\ \hline
\end{tabular}
\caption{Planet Radius and Mass are calculated using R$_{p}$/R$_{*}$ and Semi-Amplitude listed here and stellar parameters from Table \ref{tab:stars}. Semi-Amplitudes are measured in this work along with the R$_{p}$/R$_{*}$ for TOI-1011. Citations for literature R$_{p}$/R$_{*}$ Values are as follows: D=\cite{2021AJ....162...62D}, M=\cite{2016ApJ...822...86M}, P=\cite{polanski}. The value in parenthases following the last digit on Orbital Period represents the uncertainty on the last digit (example: 1.037611(9) is the same is 1.037611 $\pm$ 0.000009)}
\label{tab:planets}

\end{center}

\end{table*}

We present the RV analysis and mass measurements for 6 planets below. We then compute the CMF of each planet and compare to that of their host star where possible. The masses and CMFs for each planet are listed in Table \ref{tab:planets}. We also include Kepler-102 d (see Chapter 4) in our tables and plots (not part of the 10 planet sample we observed). A list of the host star parameters we used can be found in Table \ref{tab:stars}.

\subsection{Kepler-100 b}

Kepler-100 (KOI-41) is a Sun-like star hosting three transiting super-Earth and sub-Neptune sized planets discovered with \textit{Kepler}, and an additional outer sub-Saturn sized planet discovered with RVs \citep{kgps}. The innermost planet (Kepler-100 b) is likely rocky with an orbital period of P$_{b}$=6.89 days \citep{2019RAA....19...41G}. Kepler-100 c is likely a gaseous sub-Neptune with a radius of R$_{c}$=2.35 $\pm$ 0.20 R$_{\oplus}$ at an orbital period of P$_{c}$=12.82 days. \footnote{The citations on the planet radii and periods on planets c and d are the same as for planet b} Typically, in multi-planet systems with both gaseous and rocky planets we expect the smaller planets to be interior to the larger ones \citep{2013ApJ...763...41C, 2013ApJ...776....2L, 2018AJ....155...48W}. However, Kepler-100 d is also likely a rocky planet with a radius of R$_{d}$=1.6 $\pm$ 0.2 R$_{\oplus}$ at an orbital period of P$_{d}$=35.33 days. 

Previously, both RVs and TTVs have been used to measure the masses of the Kepler-100 planets. Initial RV measurements gave only upper limits for planets c and d, while giving a very high mass of M$_{b}$=7.3 $\pm$ 3.2 M$_{\oplus}$ for planet b \citep{2014ApJS..210...20M}, which, combined with the literature radius (R$_{b}$=1.35 $\pm$ 0.06 R$_{\oplus}$, \citealt{berger2018}), suggests an extremely iron rich planet with a CMF$_{b}$=0.97. \cite{2022AJ....163...91J} use TTVs to measure masses for all three transiting planets, demonstrating the gaseous nature of planet c (M$_{c}$=14.6 $\pm$ 2.8 M$_{\oplus}$) and giving a surprisingly low mass for planet d (M$_{d}$=1.1 $\pm$ 0.5 M$_{\oplus}$, CMF$_{d}$=-0.9). Most recently, \cite{kgps} measured RV masses for planets b (M$_{b}$=5.5 $\pm$ 1.3 M$_{\oplus}$) and c (M$_{c}$=3.8 $\pm$ 1.7 M$_{\oplus}$), and only measure a mass upper limit for planet d. This mass for planet b still suggests an extremely high iron content with a CMF$_{b}$ of 0.8, indicating that this might be an iron-enriched super-Mercury planet.  

We collected 31 RVs on this system with MAROON-X over 3 semesters, and we collected 8 RVs using HIRES. In addition, we utilized archival HIRES RVs from \cite{kgps}).

We fit the four known planets in the system, assuming zero eccentricity orbits for the three transiting planets, as is typical for compact multi-planet systems  \citep{2013ApJ...774L..15D, 2019AJ....157...61V, 2019AJ....157..198M, 2021AJ....162...55Y}. We then confirmed that fixing e=0 provided a superior fit by comparing the BIC of our preferred model to one with non-zero eccentricities. We allowed the eccentricity of the outer giant to vary, with a Gaussian prior centered on an eccentricity of 0.03 with a width of 0.1 \citep{kgps}.

\begin{figure*}
\centering
\gridline{\fig{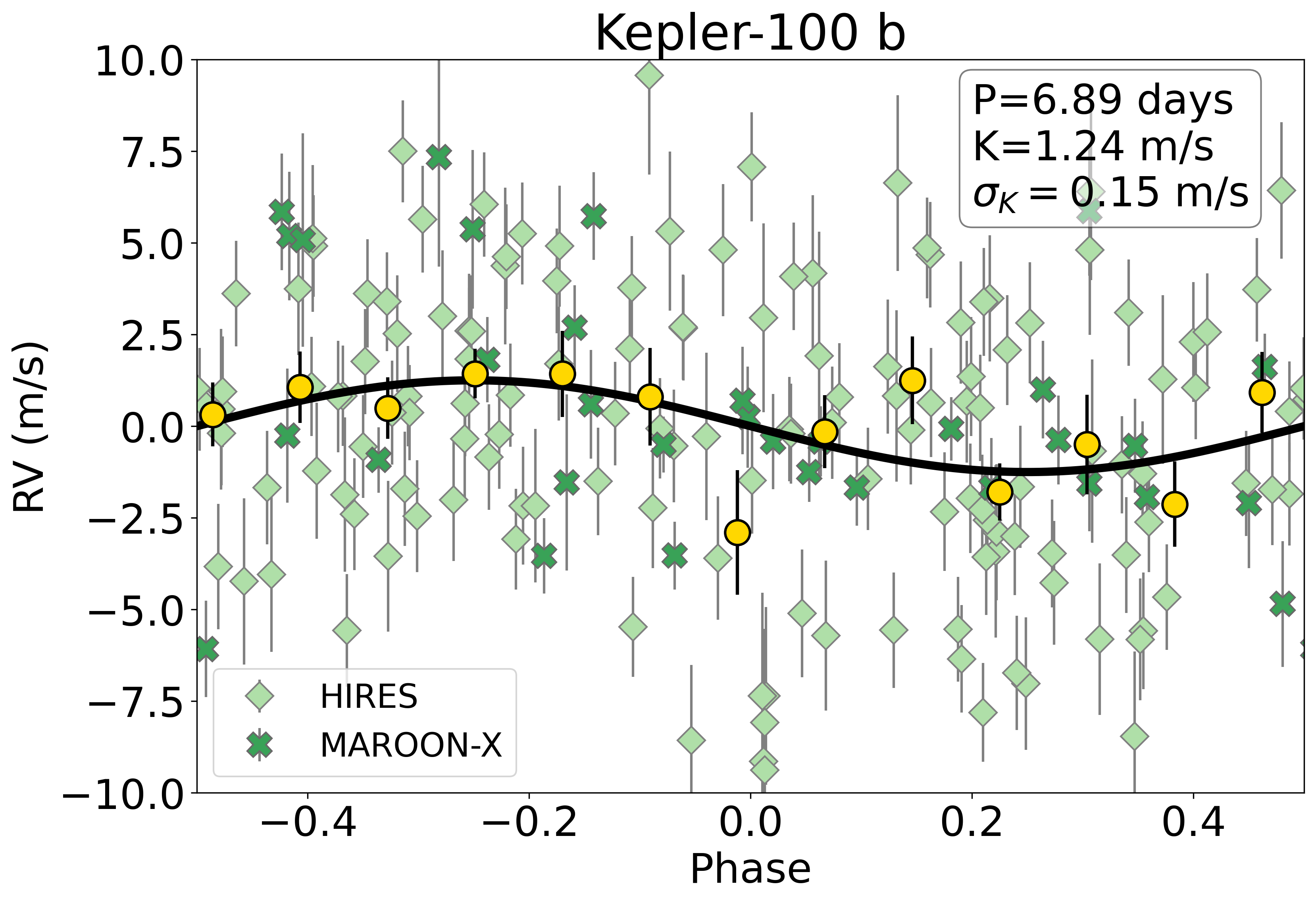}{0.49\textwidth}{}
          \fig{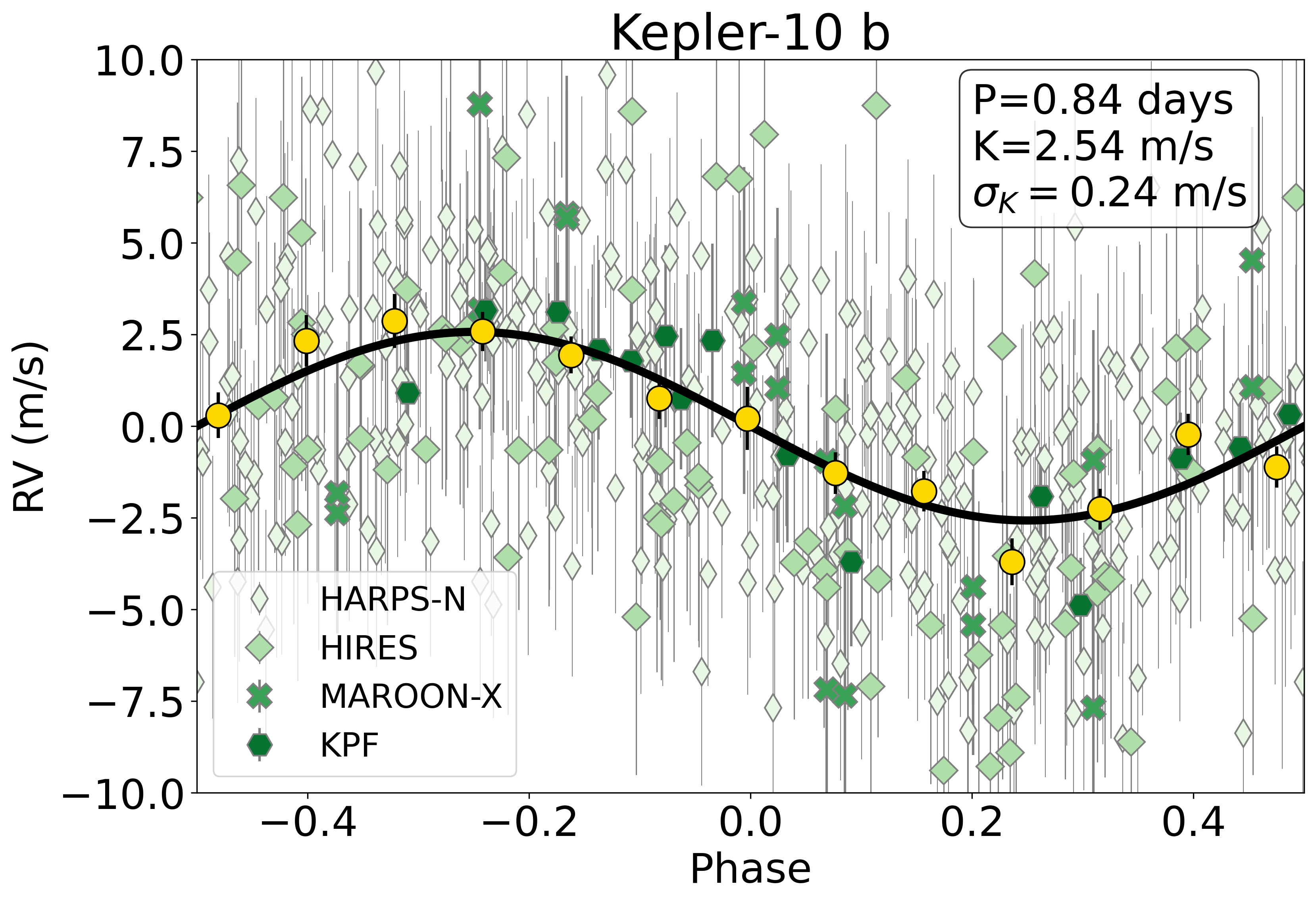}{0.49\textwidth}{}    
}
\gridline{\fig{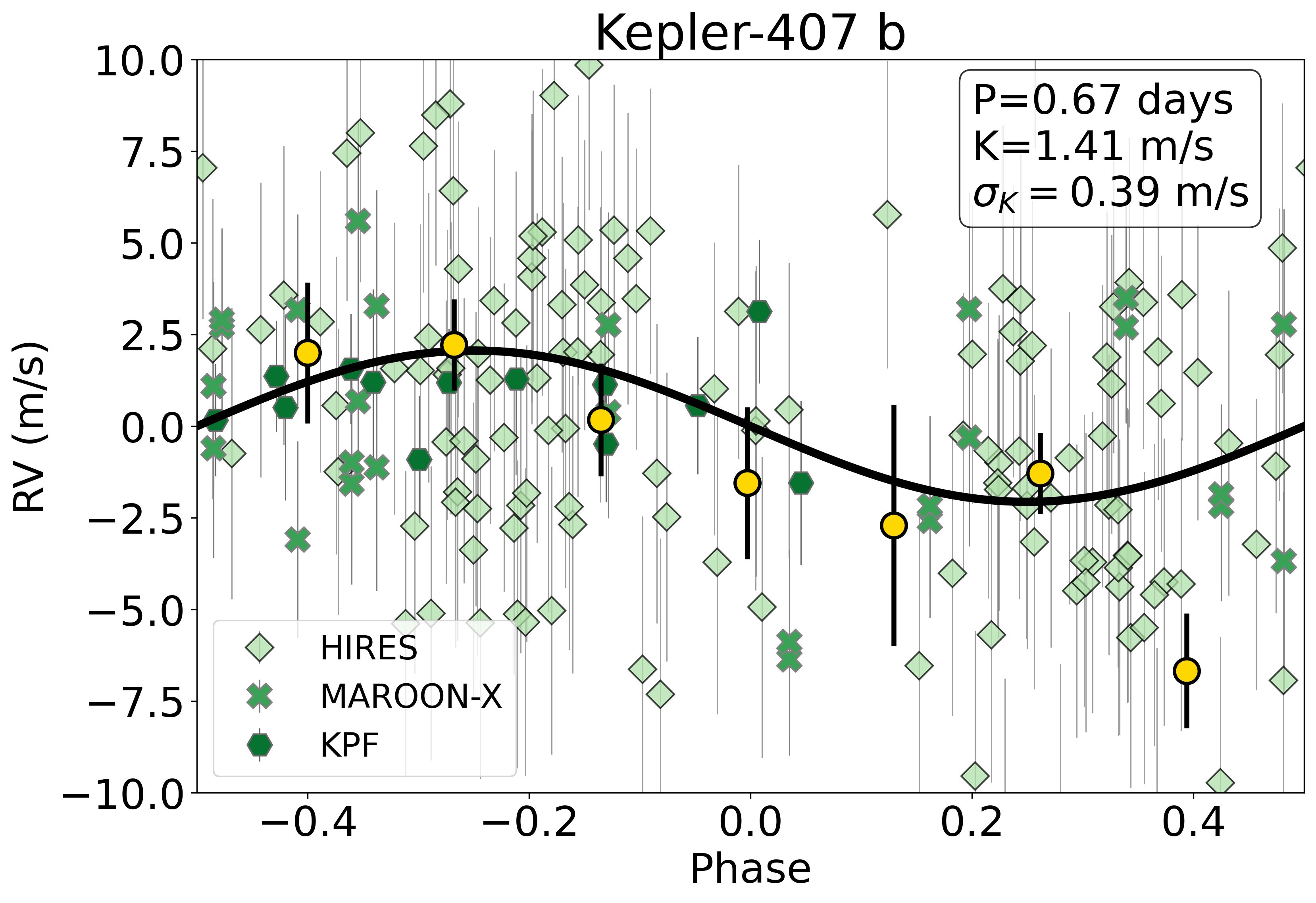}{0.49\textwidth}{}
          \fig{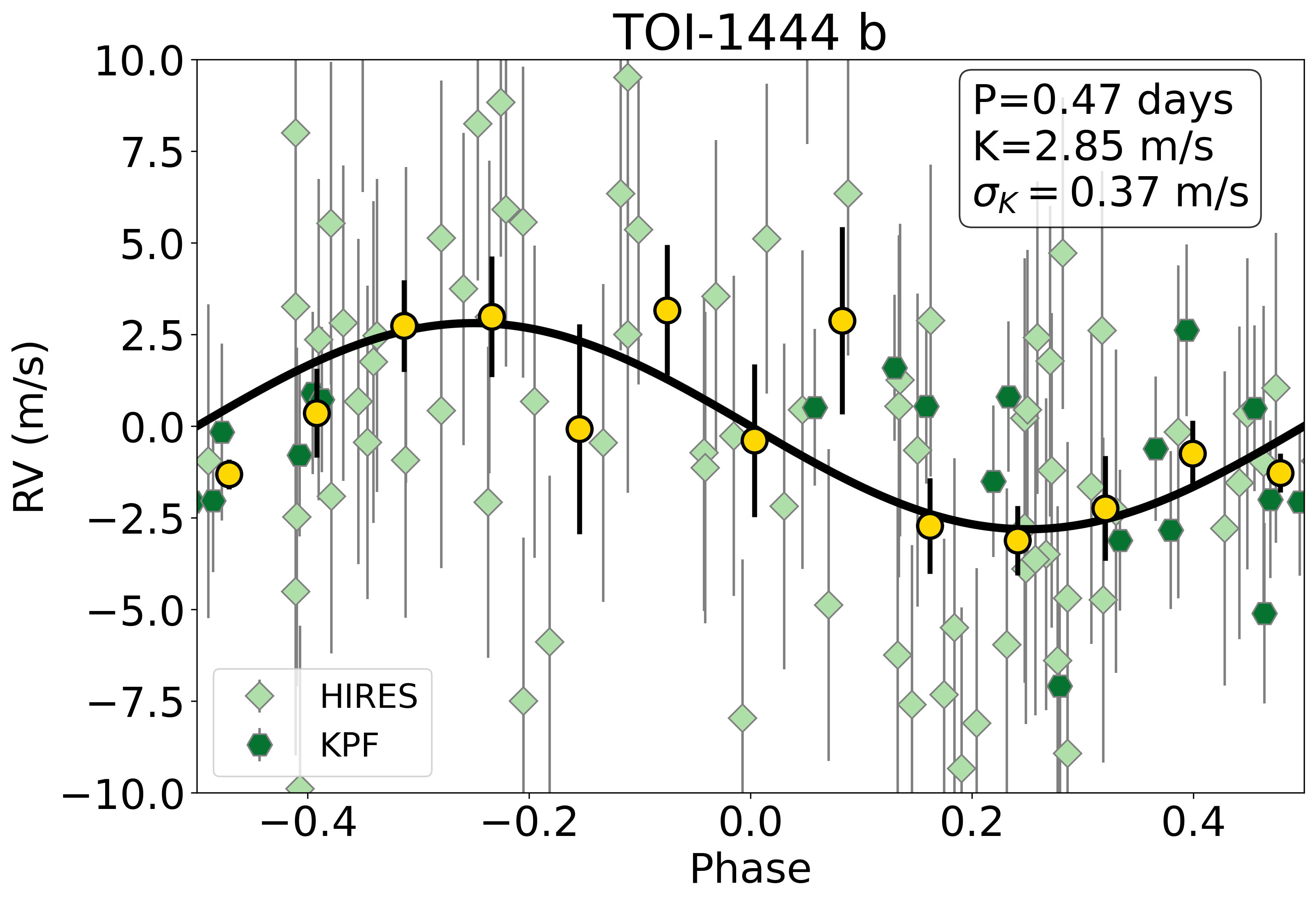}{0.49\textwidth}{}    
}
\gridline{\fig{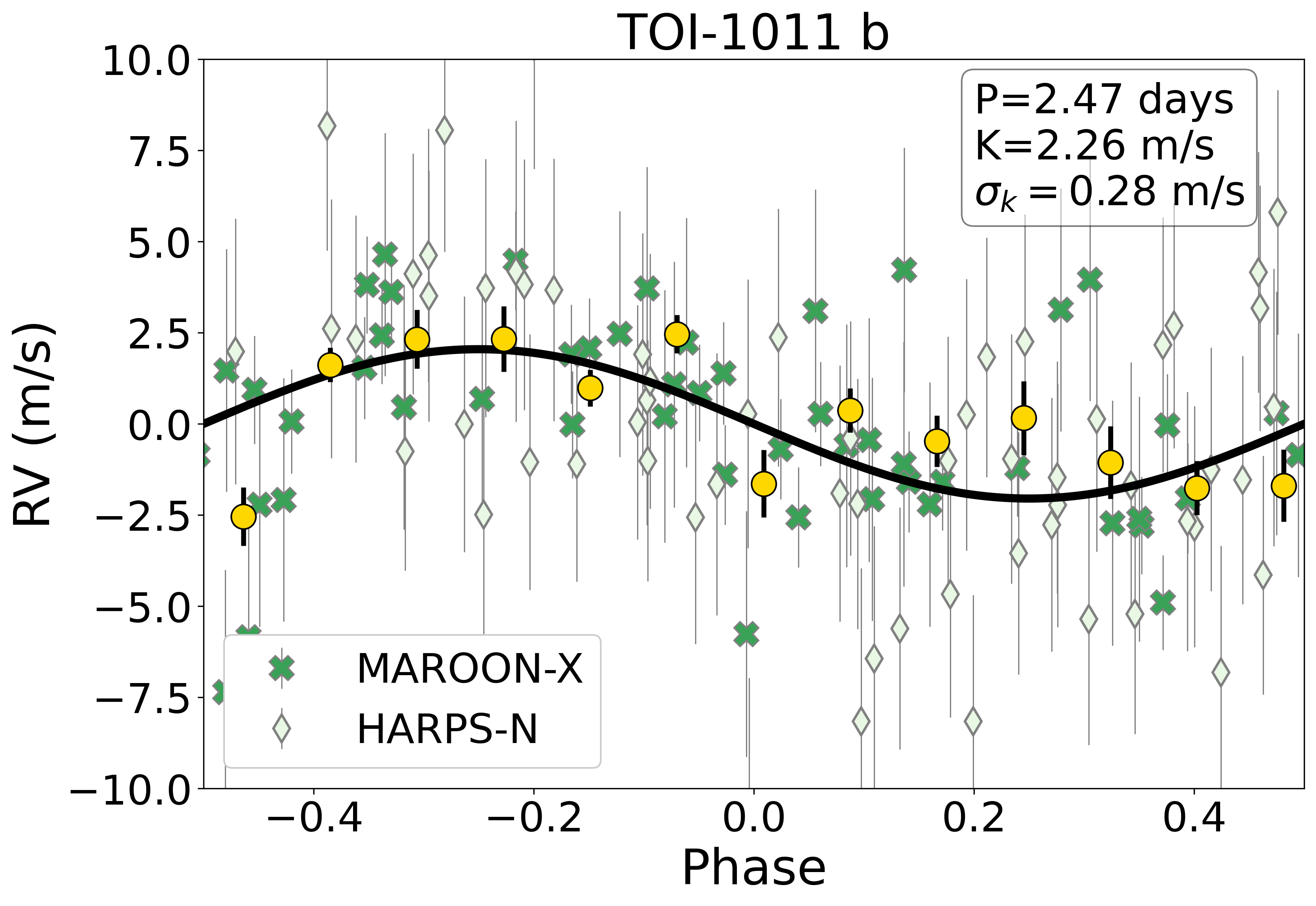}{0.49\textwidth}{}
          \fig{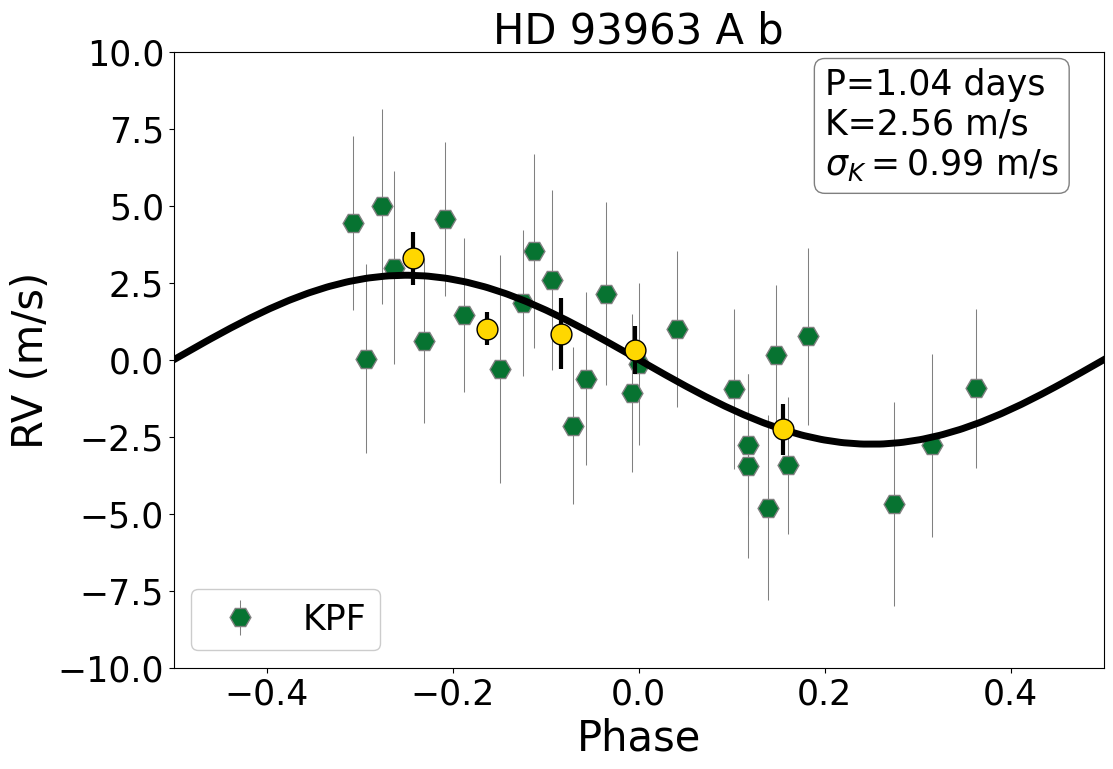}{0.49\textwidth}{}    
}
 \caption{Radial velocities phase-folded at the period of each planet, after subtracting the RV components from the other three planets based on our best-fit model. The model RV curve for each planet is overplotted in grey, with the model period, semi-amplitude (K), and the standard deviation in semi-amplitude ($\sigma$) shown. The gold points are the binned weighted median values with standard deviation of binned RVs as uncertainties.}
    \label{fig:rvs}
\end{figure*}

We measured a semi-amplitude of K$_{b}$=1.25 $\pm$ 0.15 m/s, which translates to a planet mass of M$_{b}$=3.94 $\pm$ 0.47 M$_{\oplus}$ using the updated stellar mass. With our homogeneously measured stellar radius and R$_{p}$/R$_{*}$ from \cite{2016ApJ...822...86M}, we measure an updated radius of R$_{b}$=1.34 $\pm$ 0.12 R$_{\oplus}$. Combining mass and radius measurements, we calculated a Core Mass Fraction of CMF$_{b}$=0.59 $\pm$ 0.30. The 1$\sigma$ upper bound on CMF suggests that Kepler-100 b could potentially be an iron-rich super-Mercury, but our mass suggests the planet is lower density than previously determined.

\subsection{Kepler-10 b}
Kepler-10 (KOI 72) b was the first rocky planet discovered by the \textit{Kepler} mission \citep{2011ApJ...729...27B}. It is an ultra-short period planet orbiting its host every 0.84 days. Kepler-10 also hosts two outer companions: a P$_{c}$=45.45 day sub-Neptune sized transiting planet, and a non-transiting planet with a period of P$_{d}$=151.0 days and a minimum mass of Msin(i)=12.68 M$_{\oplus}$ \citep{bonomo2023}. There is also structure in the periodogram of the RV residuals with power at P= 25 days, potentially suggestive of a planet candidate at this period \citep{kgps}, although this has previously been attributed to a harmonic of the rotational period of the star \citep{2017MNRAS.471L.125R}. 
 
As a chemically and dynamically confirmed thick-disc star \citep{2014ApJ...789..154D}, Kepler-10 has lower metallicity (-0.17 $\pm$ 0.04), and an [Fe/Mg] ratio consistent with a CMF$_{*}$ of 0.25 \citep{2024arXiv240908361B}. As such, constraining the interior composition of Kepler-10 b would be very helpful to both understand the relationship between planet and host star, and the compositions of old rocky planets around thick-disc stars. Recent RV surveys from \cite{kgps} and \cite{bonomo2023} measure a mass of 3.7 $\pm$ 0.4 M$_{\oplus}$ and 3.26 $\pm$ 0.3 M$_{\oplus}$ (respectively), suggestive of an Earth-like density but a lower CMF$_{b} \approx$ 0.15. 

To improve upon this measurement, we collected 9 RVs with MAROON-X in 2021 B, with 10 simultaneous RVs from HIRES. We also obtained 15 RVs with KPF in 2023A. In addition, we utilized the 79 previously collected HIRES RVs \citep{kgps, 2016ApJ...819...83W}, as well as 291 RVs from HARPS-N \citep{bonomo2023}. 

We fit for the three planets in the system using a simple Keplerian fit for our full dataset. We report a value of K$_{b}$=2.57 $\pm$ 0.24 m/s, which translates to a planet mass of M$_{b}$=3.58 $\pm$ 0.33 M$_{\oplus}$. With our homogeneously measured stellar radius and R$_{p}$/R$_{*}$ from \cite{2019ApJ...883...79D}, we measure an updated radius of R$_{b}$=1.47 $\pm$ 0.03 R$_{\oplus}$. Combining our mass and radius measurements, we report a CMF$_{b}$=0.16 $\pm$ 0.20. This is consistent with the CMF suggested by its host star abundances (0.25 $\pm$ 0.03).

\subsection{Kepler-407 b}

Kepler-407 b is a R$_{b}$=1.161 $\pm$ 0.039 R$_{\oplus}$ planet \citep{kgps} orbiting a G-type star on an ultra-short period orbit of P$_{b}$= 0.67 days \citep{2019RAA....19...41G}. Additionally, this system hosts a non-transiting outer giant companion (M$_{c} \approx$ 11 M$_{J}$) that sits on the boundary between giant planets and brown dwarfs with an orbital period of P$_{c}$=2096 $\pm$ 5 days \citep{kgps}. Kepler-407 is one of the most metal-rich rocky planet hosts (0.35 $\pm$ 0.05, \citealt{2018ApJS..237...38B}). In addition to being iron-rich, Kepler-407 is also rich in Magnesium ([Mg/H]=0.32 dex) with a ratio of [Fe/Mg] that would suggest a relatively Earth-like CMF$_{b}$=0.35 \citep{2018ApJS..237...38B}. 

This planet was confirmed with 17 RVs, but a precise mass measurement has proven very difficult to achieve \citep{2014ApJS..210...20M}. \cite{kgps} recently measure a mass of M$_{b}$=1.5 $\pm$ 0.9 M$_{\oplus}$ using 70 HIRES RVs. To improve this mass measurement, we collected 10 RVs with HIRES and 13 with MAROON-X in 2021 B, along with 13 RVs from KPF in 2023 A. 

We fit these RVs using a three planet model, and recover a best-fit semi-amplitude of K$_{b}$=1.41 $\pm$ 0.39 m/s (Figure \ref{fig:rvs}). We get consistent solutions when fitting each RV set individually (HIRES: 1.63 $\pm$ 0.41, KPF: 1.72 $\pm$ 1.1, MAROON-X:1.3 $\pm$ 0.69). Using our best-fit K$_{b}$ and our updated stellar mass, we recover a mass of M$_{b}$=1.93 $\pm$ 0.50 M$_{\oplus}$. 

With our homogeneously measured stellar radius and R$_{p}$/R$_{*}$ from \cite{2016ApJ...822...86M}, we measure an updated radius of R$_{b}$=1.19 $\pm$ 0.05 R$_{\oplus}$. The mass and radius measurements for Kepler-407 b suggest an Earth-like composition of CMF$_{b}$=0.35 $\pm$ 0.32, which is also consistent with that of its host star (CMF$_{*}$=0.40). 

\subsection{TOI-1444}

TOI-1444 is solar metallicity ([Fe/H]=0.03) G-type star hosting an ultra-short period (0.47 days) rocky planet with a radius of $\sim$1.4 R$_{\oplus}$. The rocky USP has an outer, non-transiting companion with an Msin(i) consistent with a sub-Neptune-sized planet \citep{2021AJ....162...62D}. Previously, \cite{2021AJ....162...62D} measured a semi-amplitude for the rocky USP of K$_{b}$=3.30 $\pm$ 0.59 m/s and a mass of 3.78 $\pm$ 0.71 M$_{\oplus}$. To measure a more precise mass, we obtained 18 RVs from KPF in August 2023. 

Using a two planet model we measure a best-fit semi-amplitude of K$_{b}$=2.85 $\pm$ 0.37 m/s using both the HIRES and KPF datasets (Figure \ref{fig:rvs}).  Individually, the HIRES and KPF datasets produce values for K$_{b}$ of 3.2 $\pm$ 0.4 m/s  and 3.3 $\pm$ 0.4 m/s, respectively. This semi-amplitude produces a planet mass of M$_{b}$=3.34 $\pm$ 0.43 M$_{\oplus}$. With our homogeneously measured stellar radius and R$_{p}$/R$_{*}$ from \cite{polanski}, we measure an updated radius of R$_{b}$=1.42 $\pm$ 0.04 R$_{\oplus}$. These masses and radii give a CMF of CMF$_{b}$=0.22 $\pm$ 0.17, which is consistent with an Earth-like composition and that of its host star (CMF$_{*}$=0.29 $\pm$ 0.03) to within 1$\sigma$. 

\subsection{TOI-1011 b}
\label{sec:toi1011}
The TESS mission \citep{Ricker2015} collected 2-minute cadence photometry of TOI-1011 (HD 61051, TIC 114018671) in Sectors 34 and 61 (January 2021 and January 2023, respectively). TOI-1011 is a previously unconfirmed planet flagged in both SPOC and QLP pipelines in sector 34 photometry \citep{2020RNAAS...4..201C, 2022ApJS..259...33K}. The parameters from these pipelines suggest an orbital period of P=2.4705 $\pm$ 0.0000073 days and a planet radius of R$_{b}$=1.45 $\pm$ 0.11 R$_{\oplus}$. We used Archival HARPS-N RVs \citep{2020A&A...636A..74T} to identify a signal with a semi-amplitude of K$_{b}$=2.4 $\pm$ 0.8 m/s for a single planet model at the orbital period of the planet candidate. 

To measure a precise mass for the planet, we obtained 40 RVs with MAROON-X in 2022 A and 2023 B. Using a single planet model we measure a best-fit semi-amplitude of K$_{b}$=2.26 $\pm$ 0.28 m/s. Using the stellar mass measured in Section \ref{sec:starmass} we calculate a planet mass of M$_{b}$=4.59 $\pm$ 0.56 m/s. This confirms the planetary nature of the signal with a significance of 8$\sigma$. 

\begin{figure}
    \centering
    \includegraphics[width=0.5\textwidth]{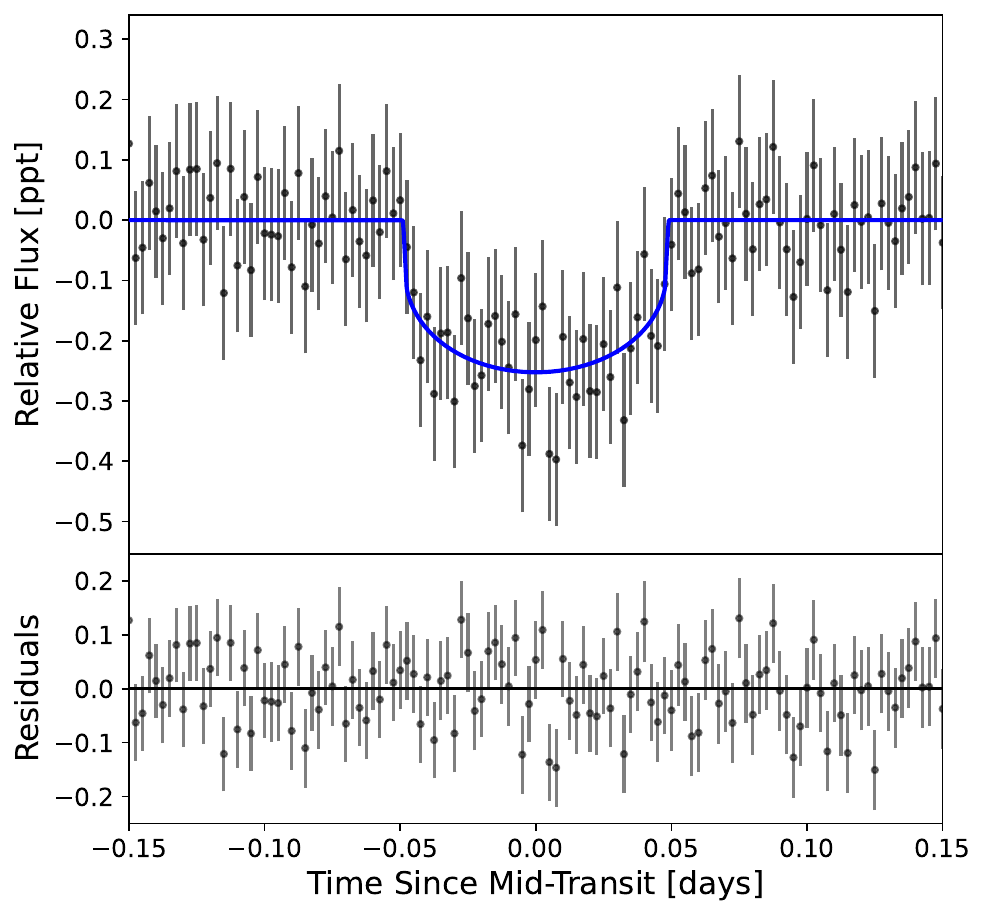}
    \caption{Phase-folded transit light curve of TOI-1011 b from sectors 34 and 61 of \textit{TESS} photometry. We have binned the phased light curve (black) and show the best-fit transit model in blue. The bottom panel shows the binned residuals after removing the best fit model. }
    \label{fig:TOI1011}
\end{figure}




To measure the planet radius we used the Presearch Data Conditioning Simple Aperature Photometry (PDC-SAP) light curves \citep{Stumpe2012,Stumpe2014} produced by the TESS Science Processing Operations Center \citep[SPOC;][]{2016SPIE.9913E..3EJ}. We removed long-term variability and systematics from the light curves using an iterative sigma-clipping spline fit \citep[using {\tt wōtan};][] {2019AJ....158..143H}. We first masked the transits according to the orbital period and time of inferior conjunction of TOI-1011 b from the SPOC pipeline on sector 34 photometry, then applied the detrending routine to the transit-masked light curves to ensure the transit signal was not removed. 

We employed {\tt exoplanet} \citep{2021JOSS....6.3285F} and {\tt pyMC3} \citep{2016ascl.soft10016S} to construct the transit model for TOI-1011 b. Transit parameter priors (depth, duration, orbital period, and time of conjunction)\footnote{we assumed a circular orbit ($e=0$) in our transit fit similarly to our RV fit.} were adopted from the SPOC pipeline on Sector 34 data, and stellar parameter priors (radius and density) from our isochrone fits. The priors (Table \ref{tab:1011trans}) were uniformly or normally distributed with large sigmas to allow for a thorough and unconstrained search of the parameter space. We supplied the stellar density prior to mitigate the degeneracy between the impact parameter $b$ and scaled semi-major axis $a$/$R_\star$ of the planet. Limb darkening coefficients q$_1$ and q$_2$ were adopted following the quadratic parameterization of \citet{Kipping2013}. The transit model further included the transit duration and the planet-to-star radius ratio R$_{\rm p}$/R$_{\star}$, derived from the transit depth. We fit all the transits assuming a linear ephemeris as we did not identify any significant transit timing variations. The initial constant-period maximum-likelihood model served as an initial guess for posterior sampling. We then performed a MCMC analysis to sample the posterior distributions of the transit parameters. We find R$_{\rm p}$/R$_{\star}$=0.0143 $\pm$ 0.0006 which translates to a planet radius of R$_{b}$=1.45 $\pm$ 0.05 R$_{\oplus}$.  

Combining our radius and mass measurements, we compute a CMF$_{b}$=0.33 $\pm$ 0.26. This suggests TOI-1011 is a planet with an Earth-like composition. 

\begin{deluxetable*}{lccc}
    \label{tab:1011trans}
    \tabletypesize{\scriptsize}
    \tablecaption{Transit Parameters of TOI-1011 b}
    \tablehead{\colhead{Parameter} & \colhead{Symbol} & Prior & \colhead{Posterior (Median and 68.3\% CI)} }
    \startdata
    Orbital Period (days) & P$_{\rm orb}$ & $\mathcal{N}$ (2.4697, 10) & 2.470498 $\pm$ 0.000007\\
    Time of Conjunction (BJD-2457000) & T$_c$ & $\mathcal{N}$ (2213.1309, 10) & 1489.97760 $\pm$ 0.00383\\
    Transit Duration (hrs) & & $\mathcal{N}$ (2.3244, 10) & 2.3016 $\pm$ 0.0341\\
    Eccentricity & $e$ & 0 (fixed) & 0  \\
    Impact Parameter$^\dagger$ & $b$ & $U$ (0, 1+R$_{\rm p}$/R$_{\star}$) & 0.24 $\pm$ 0.15 \\
    Limb Darkening q$_1$ &  & $U$ (0, 1) & 0.55 $\pm$ 0.36 \\
    Limb Darkening q$_2$ &  & $U$ (0, 1) & 0.09 $\pm$ 0.35 \\
    Planet/Star Radius Ratio & R$_{\rm p}$/R$_{\star}$ & $\mathcal{N}\propto$ transit depth & 0.0143 $\pm$ 0.0006\\
    \enddata
\end{deluxetable*}

\begin{figure}
    \centering
    \includegraphics[width=8cm]{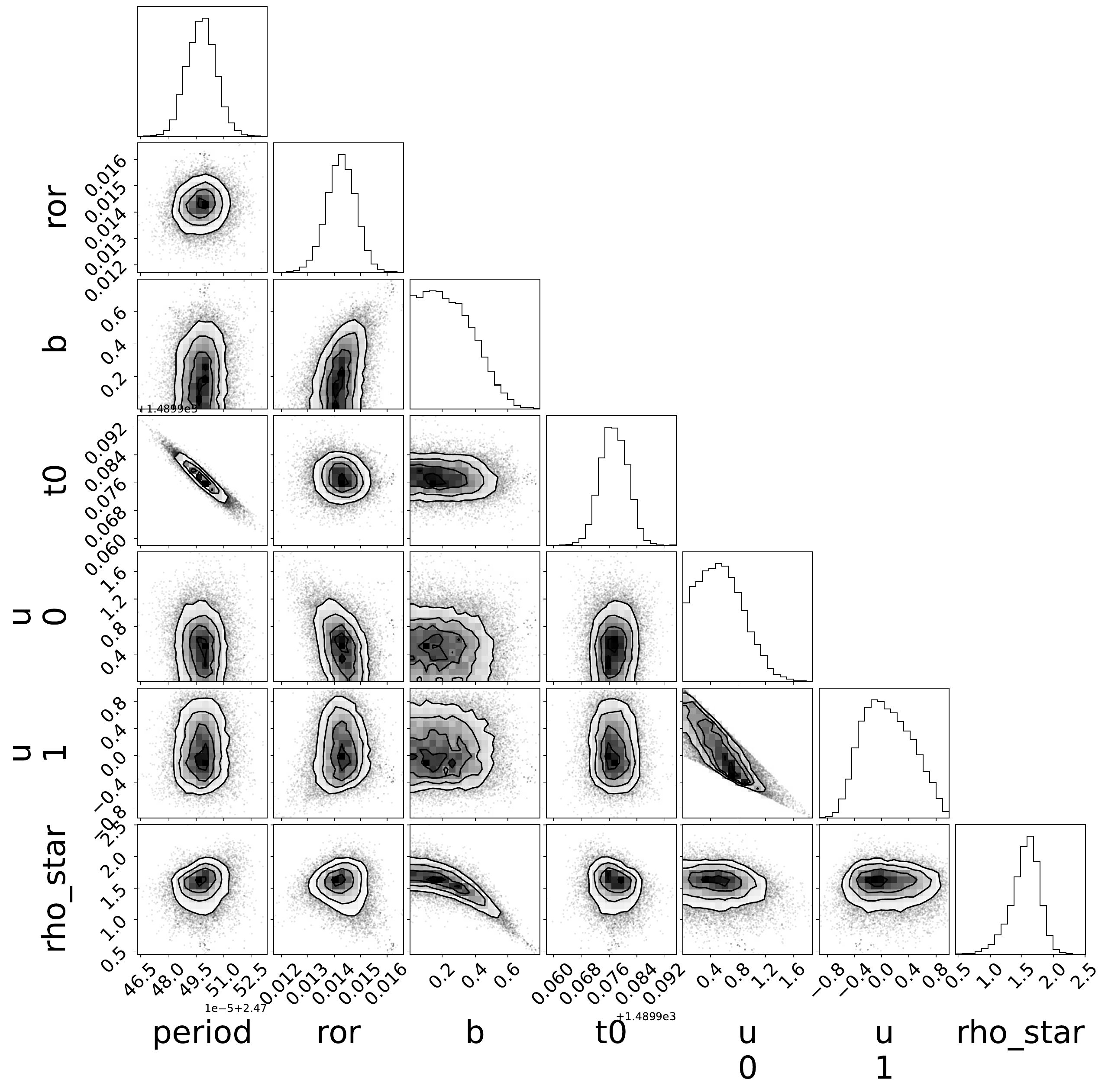}
    \caption{Corner plot of key transit model fit parameters for TOI-1011 b.}
    \label{fig:corner}
\end{figure}


\begin{figure*}
    \centering
    \includegraphics[width=1.0\textwidth]{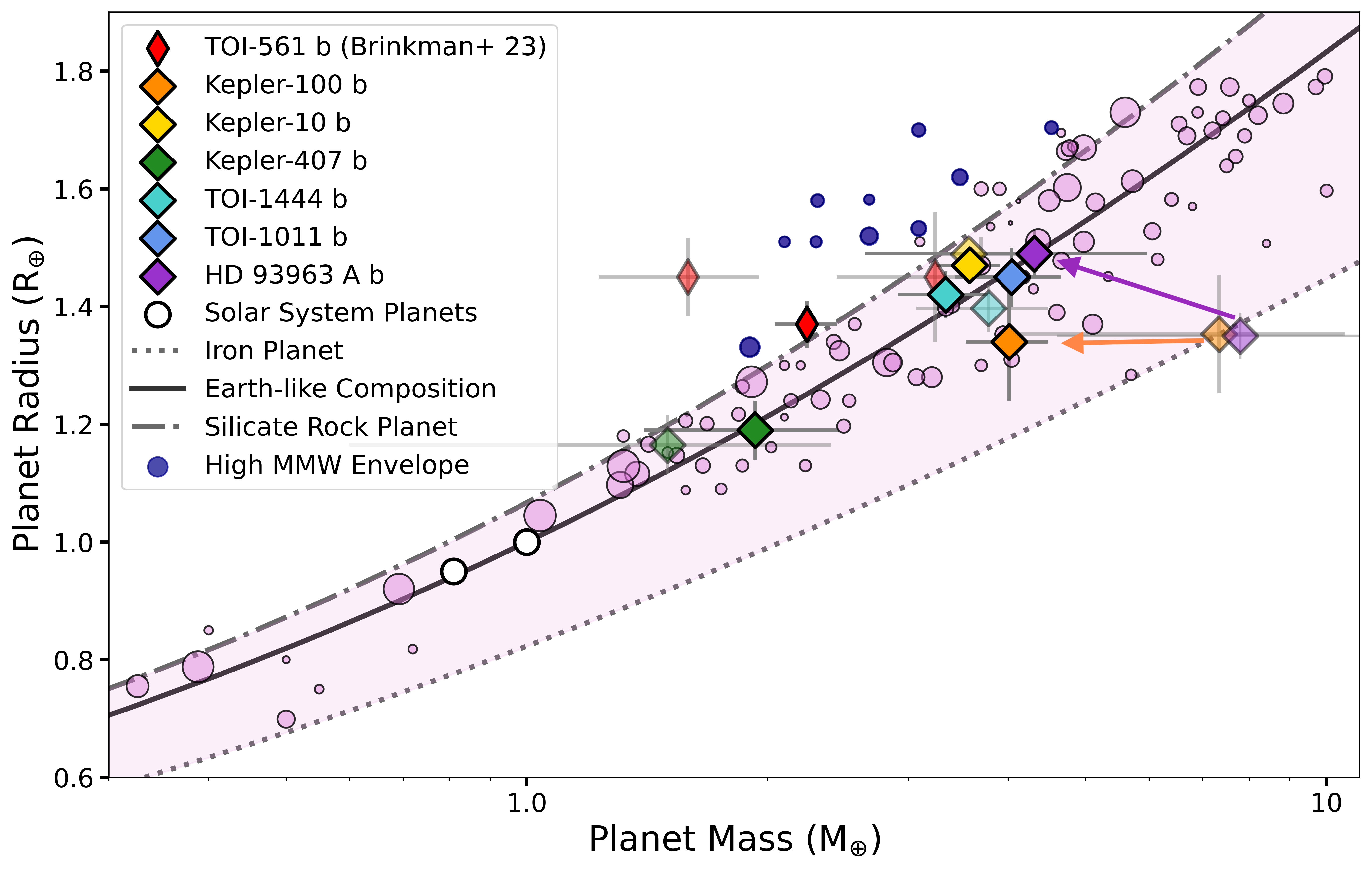}
    \caption{Planet radius vs. mass for our sample (diamonds) are shown in the context of the rocky planet population (circles). Opaque diamonds represent our mass and radius measurements while translucent diamonds are the previously published values. Lines show mass-radius relations for planets with an Earth-like composition, a purely silicate-rock composition, and a purely iron composition \citep{2019PNAS..116.9723Z}. Literature values come from the NASA Exoplanet Archive  \citep{2013PASP..125..989A} queried 4/12/2024 for planets with mass and radius measurements with fractional uncertainties $<$50$\%$, and the circle size is inversely correlated with fractional uncertainty in density. Planets consistent with a rocky composition (mixture of iron and silicate only) are shown in pink. Planets with masses and radii 1$\sigma$ away from a pure silicate composition (dashed line) are shown in purple, and possibly host envelopes made of high-mean molecular weight (MMW) species.}
    \label{fig:massrad}
\end{figure*}

\begin{figure*}
    \centering
    \includegraphics[width=1.0\textwidth]{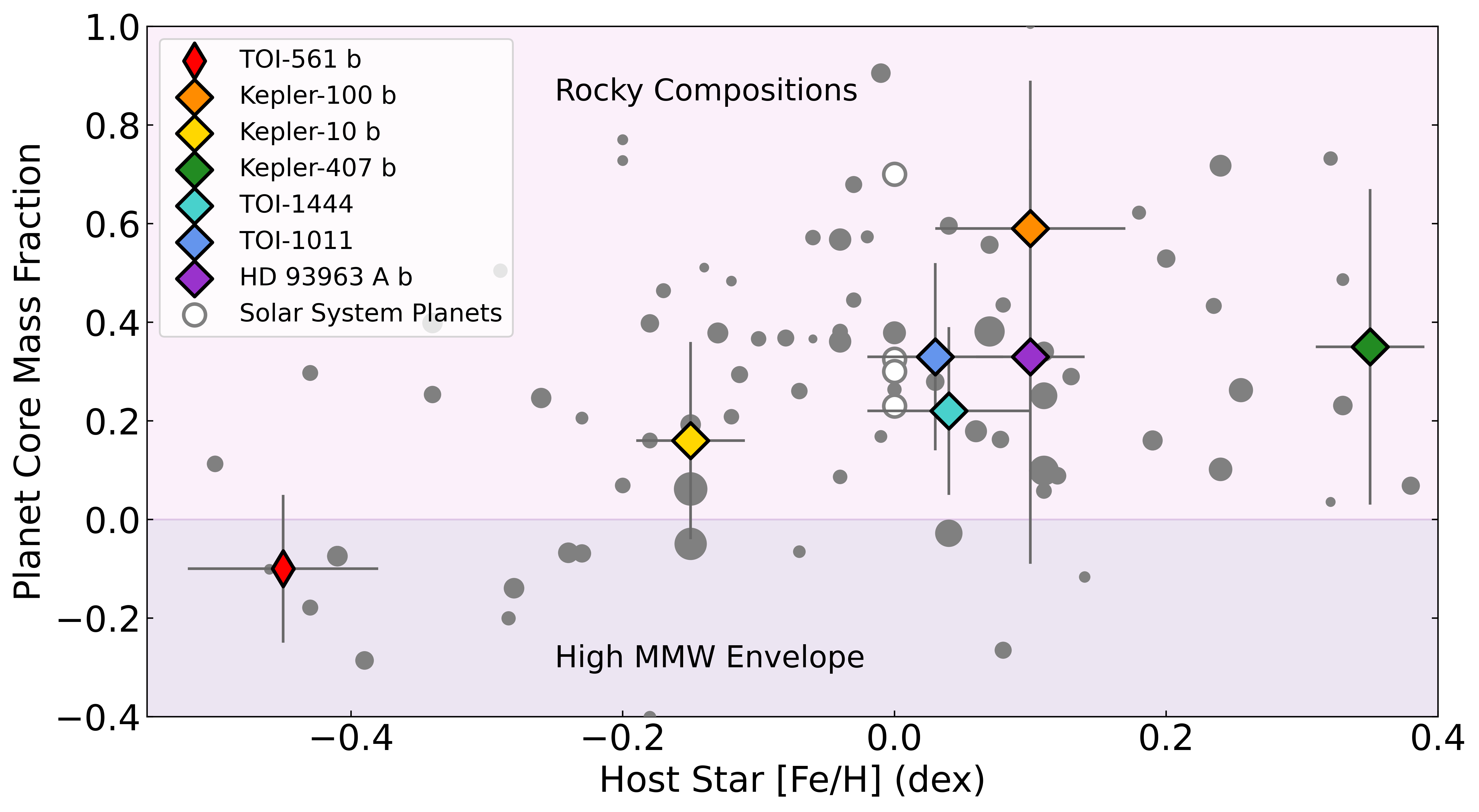}
    \caption{The Core Mass Fraction (CMF) of rocky exoplanets vs. [Fe/H] metallicity for their host stars is shown for our sample in the context of the sample of rocky planets. The colors of each diamond correspond to the same planet as listed in Figure \ref{fig:massrad}. The rocky planet point size is inversely correlated with fractional uncertainty in CMF.}
    \label{fig:metallicity}
\end{figure*}

\begin{figure*}
    \centering
    \includegraphics[width=1.0\textwidth]{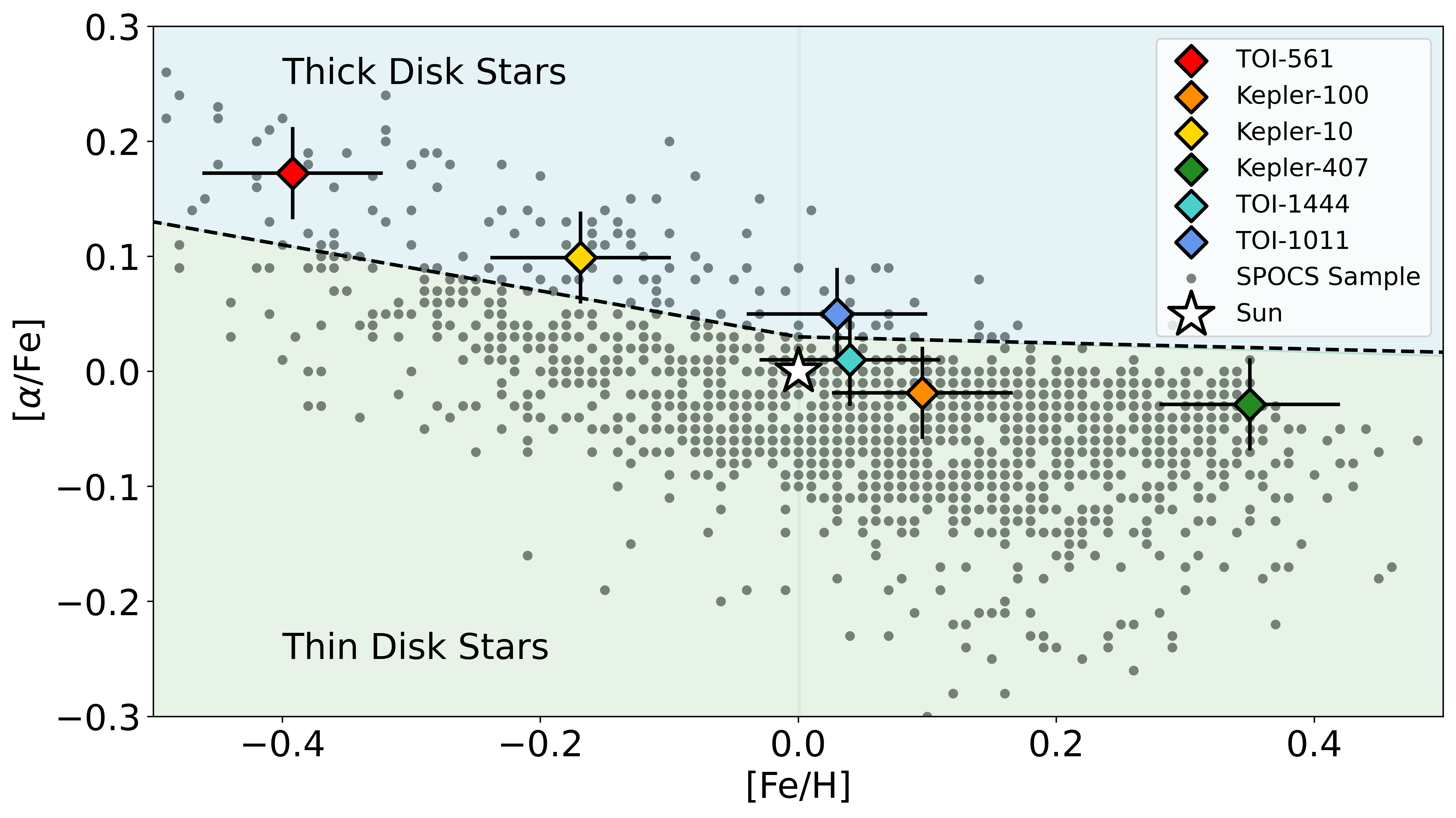}
    \caption{Alpha abundances [$\alpha$/Fe] as a function of [Fe/H] for each host star in our sample, along with a larger sample of exoplanet hosts from the SPOCS catalogue \citep{2018ApJS..237...38B}. The colors of each diamond correspond to the same planet as listed in Figure \ref{fig:massrad}. The black dashed line approximately separates the galactic thin disc and thick disc populations \citep{2021AJ....161...56W}. The two stars in our sample that fall above this line and were likely born in the galactic thick disc are labeled.}
    \label{fig:alpha}
\end{figure*}

\begin{figure*}
    \centering
    \includegraphics[width=1.0\textwidth]{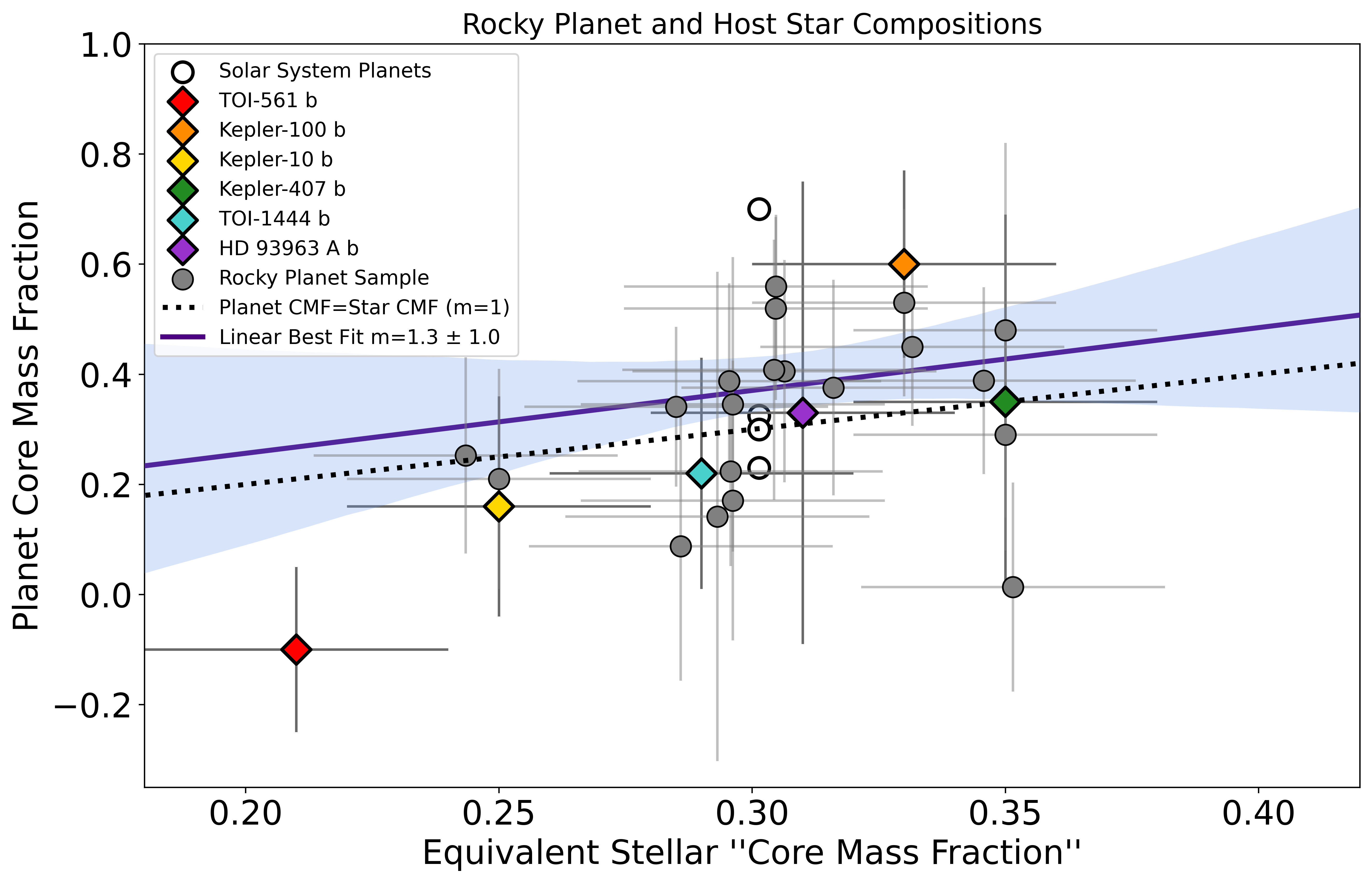}
    \caption{The Core Mass Fraction (CMF) of rocky exoplanets vs equivalent CMF for their host stars is shown for our sample in the context of the sample of rocky planets (from \cite{2024arXiv240908361B}). The CMF for the four inner solar system planets are shown at the equivalent CMF of the Sun (from top to bottom: Mercury, Earth, Venus, and Mars). The dotted line shows the one-to-one correspondence of CMF in stars and planets, which is where planets would fall if they inherited the exact Fe/Mg ratio from their host star. Our linear best fit model is shown in indigo (Ordinary Least Squares with rejection sampling.}
    \label{fig:cmf}
\end{figure*}

\subsection{HD 93963 A b}
HD 93963 A b is a planet in an S-type orbit around a binary star, with an outer transiting Neptune-sized companion (HD 93963 A c). This rocky planet was previously reported to have a mass of M$_{b}$=7.8 $\pm$ 3.2 M$_{\oplus}$ and a radius of R$_{b}$1.35 $\pm$ 0.04 R$_{\oplus}$, which would suggest an extremely iron rich planet with a CMF$_{b} >$ 1 \citep{2022A&A...667A...1S}.  We collected 30 RVs on HD 93963 A with KPF over 9 nights from November 2023-January 2024. 


We fit our full dataset using a two planet model, and measure a best-fit semi-amplitude is K$_{b}$= 2.56 $\pm$ 0.99 m/s for the full dataset (Figure \ref{fig:rvs}). We get similar values using only the KPF RVs (K$_{b}$=2.62 $\pm$ 1.3 m/s). Using the stellar mass measured here we calculate a planet mass of M$_{b}$=4.31 $\pm$ 1.66 M$_{\oplus}$. 

Recently \cite{polanski} measured an updated value for R$_{p}$/R$_{*}$ that is larger than the value from \cite{2022A&A...667A...1S} by 2$\sigma$ (0.0131 $\pm$ 0.0006, compared to 0.01190 $\pm$ 0.00036). We adopt the \cite{polanski} value and use the stellar radius measured here to produce an updated radius of R$_{b}$=1.49 $\pm$ 0.042 R$_{\oplus}$. 

Combining our updated values for mass and radius we calculate a Core Mass Fraction of CMF$_{b}$=0.33 $\pm$ 0.42. The uncertainty on this mass measurement (and corresponding CMF) is large, and is still therefore in agreement with a Mercury-like composition at the upper end of its 1 $\sigma$ posterior distribution. However, the median of the posterior suggests a composition more similar to that of the Earth that is consistent with its host star (CMF$_{*}$=0.31).


\section{Discussion}

\subsection{The Diversity of Rocky Planet Compositions}

Figure \ref{fig:massrad} shows our newly measured masses and radii in comparison to the previous literature values in a Mass Radius diagram. We see that for every planet (except TOI-561 b), the new mass measurement places the planet closer to an Earth-like composition (black solid line) than previously determined. Planets with a previous CMF smaller than the Earth (Kepler-10 b and Kepler-407 b) now have a larger CMF after our analysis, while planets with previous CMF larger than Earth (Kepler-100 b, HD 93963 A b, and TOI-1444 b) now appear to have a smaller CMF than previously determined. TOI-1011 b did not have a prior mass or radius measurement, but ours suggests it has an Earth-like composition (CMF=0.33). 

We quantify this observation by computing the fractional Root Mean Square Error (RMSE) of our population of planets before and after our analysis against the Earth-like composition mass-radius model \citep{2019PNAS..116.9723Z}. The fractional RMSE is the square root of mean of squared fractional residuals:
\begin{equation}
    \text{RMSE} = \sqrt{\frac{1}{n} \sum_{i=1}^{n} \left(\frac{y_p - y_a}{y_p}\right)^2}
\end{equation}
Where y$_{p}$ is the predicted radius of a planet (at a given mass) with an Earth-like composition, and y$_{a}$ is the actual observed radius of the planet at a given mass. We find an RMSE of 0.14 (14$\%$) for our planets using their initial mass measurements, and an RMSE of 0.06 after our analysis. This supports the observation that the planets in our sample appear more ``Earth-like'' after updating their masses with precision RVs. 

This is most apparent in the two high-density ``super-Mercuries'' in our sample (Kepler-100 b and HD 93963 A b) both had previous mass measurements that suggested a CMF close to 1.0 (or 100$\%$ iron). The updated mass measurements produce much lower CMFs of 0.53 for Kepler-100 b and 0.65 for HD 93963 A b. This supports studies finding that planets previously thought to be iron-rich super-Mercuries becoming more Earth-like with additional high-precision RVs \citep{2023AJ....165...97R, 2024MNRAS.529..141G}, Gaussian Process modeling to account for stellar noise \citep{Brinkman23A} or updated stellar parameters \citep{2024arXiv240908361B}. 

\subsection{Planet Host Star Connections}

Figure \ref{fig:metallicity} shows our new planet Core Mass Fractions as a function of host star [Fe/H] metallicity. We observe that planets orbiting low-metallicity stars tend to have small CMFs (TOI-561 b and Kepler-10 b), while planets orbiting high-metallicity stars have a wider range in CMF, consistent with \cite{2024arXiv240908361B}. However, much of the visual trend seen in Figure \ref{fig:metallicity} is driven by the low CMF of TOI-561 b. 

For five of the host stars in our sample (TOI-561, TOI-1444, Kepler-10, Kepler-100, and Kepler-407) we have [Mg/H] and [Fe/H] abundances (Kepler-407 from \cite{2018ApJS..237...38B}, the rest from Polanski et al. \textit{in Prep}). The abundances for these stars are shown in Figure \ref{fig:alpha} against the broader population of exoplanet host stars. 

Two stars in our sample (TOI-561 and Kepler-10) have higher $\alpha$ abundances relative to iron that are characteristic of the chemically defined thick disc \citep{2006MNRAS.367.1329R} (Figure \ref{fig:alpha}), in agreement with previous findings (\citealt{2014ApJ...789..154D, 2021AJ....161...56W, 2024arXiv240908361B}). Both TOI-561 b and Kepler-10 b have the smallest CMFs in our sample, which along with the thick disc planet HD 136352 b (CMF=0.25, \citealt{2024arXiv240908361B}) suggests that many of rocky planets orbiting thick disc stars could be iron-poor themselves instead. 

We would expect smaller CMFs around thick-disc rocky worlds if they inherit the same ratio of iron to $\alpha$ abundances (such as Mg and Si) in their host star \citep{Dorn2015}. Beyond iron to silicate ratios, it has been suggested that the planetary building blocks of thick disc stars should be more water-rich than those of the thin disc \citep{2023A&A...673A.117C}, which could contribute to the formation of secondary water envelopes. The small sample size that plagues much of this analysis is very apparent here, and the discovery and characterization of additional rocky planets around thick disc stars will illuminate whether they tend to be iron-poor or host gaseous envelopes of high mean molecular weight. 

Figure \ref{fig:cmf} shows the updated CMF of these planets versus the equivalent CMF of their host stars (as described in Section \ref{sec:starcmf}). Two of these planets (Kepler-100 b and TOI-561 b) have a CMF that deviates from their host star by $>$1$\sigma$, while the remaining three have a CMF consistent with their host stars. This is in agreement with the rocky planet sample in Figure \ref{fig:cmf}, where 75$\%$ of planets have a CMF within 1$\sigma$ of their host star. 

To quantify the relationship between planet and host star, we perform a linear fit (form $y=mx+b$) where $y$ is planet CMF, $x$ is host star equivalent CMF, $m$ is the slope, and $b$ is the intercept. We use the \texttt{curve\_fit} ordinary least squares (OLS) functionality in {\tt SciPy} to find the best-fit slope and intercept. We then use a Monte Carlo approach to estimate the uncertainties on these parameters by drawing values for CMF for each planet and star from Gaussian distributions and repeating the fit. As part of this process, we reject all values for planet density that produce a negative CMF (as this would no longer be a rocky planet) as to not bias our fit toward nonphysical values. We find a best-fit slope of m=1.3 $\pm$ 1.0 (indigo line in Fig \ref{fig:cmf}). This does not demonstrate a steep and statistically significant relationship between star and planet compositions, in agreement with \cite{2024arXiv240908361B}. 
\begin{deluxetable}{lcc}
    \label{tab:models}
    \tabletypesize{\scriptsize}
    \tablecaption{Comparison of Planet-Star Composition Relationships}
    \tablehead{\colhead{Publication} & \colhead{Fitting Method} & \colhead{Slope)} }
    \startdata
    \cite{Adibekyan2021}                 & ODR              &  6.3 $\pm$ 1.2 \\
    \cite{liu}                           & ODR              & 10.80 $\pm$ 3.56\\
    \citep{2024arXiv240908361B} & ODR              & 5.6 $\pm$ 1.6\\
    This work                            & ODR w/ TOI-561 b & 3.9 $\pm$ 1.1\\
    This work                            & ODR w/o TOI-561 b & 2.6 $\pm$ 1.2\\
    This work (best-fit)                 & OLS w/ Rejection Sampling & 1.3 $\pm$ 1.0\\
    \enddata
\end{deluxetable}

We perform two additional fits using Orthogonal Distance Regression (ODR) to better to compare our results with previous studies (namely \cite{Adibekyan2021}, \cite{liu}, and \cite{2024arXiv240908361B}). We first perform this fit using the same sample of planets as above, which includes the negative CMF value for TOI-561 b, and recover a slope of $m=3.9$ $\pm$ 1.1. However, the negative value for CMF we report for TOI-561 b is not an accurate measurement of iron to rock, and is simply an indication that it likely hosts a gaseous envelope. When we exclude TOI-561 b from the fit, and only use planets with a positive CMF, our ODR fit produces a slope of $m=2.6$ $\pm$ 1.2. This slope is only 1$\sigma$ larger than a slope of 1, and is smaller than that from  and \cite{Adibekyan2021},  \cite{liu}, and \cite{2024arXiv240908361B} (Table \ref{tab:models}).





\section{Conclusions}

We collected high-precision RVs on 6 planets using Keck/HIRES, Keck/KPF, and Gemini/MAROON-X, and report updated mass and radius measurements for each planet here, including confirmation of the planet TOI-1011 b. We then calculated the Core Mass Fractions (CMF) and compared the compositions of planet to host star where possible. Our primary conclusions are as follows: 

\begin{itemize}
    \item The CMF of all planets in our sample become closer to that of the Earth after updating their masses with precise RVs. This suggests that most rocky planets on close-in orbits have compositions similar to Earth.
    \item Two planets in our sample (Kepler-100 b and HD 93963 A b) had a previous mass and radius measurements that suggested an extremely iron-rich CMF$\approx$1. The updated mass measurements for both planets shrink significantly, and suggest compositions with smaller iron fractions (CMF of 0.53 and 0.65, respectively). 
    \item The one planet whose composition moved away from that of the Earth is TOI-561 b, which appears to be in a class of super-Earth sized planets hosting envelopes of high-mean molecular weight. 
    \item Using only planets with positive CMF values, the relationship in the iron to magnesium ratios between stars and planets is consistent with being 1-to-1. This does not support previous conclusions stating a steep and statistically significant correlation between planet and host star compositions. 
    \item Planets orbiting $\alpha$/Fe enriched thick-disc stars have the smallest CMFs in our sample when compared to more metal-rich thin-disc stars. This suggests that these planets are potentially more likely to be iron-poor or host gaseous envelopes made of high mean molecular weight species.

\end{itemize}

The diversity of Earth-sized planets in many ways is less than it once appeared to be, with the reduction of planets belonging to the iron-rich ``Super-Mercury'' population. However, planets such as TOI-561 b have radii less than 1.5 R$_{\oplus}$, yet likely host a gaseous envelope made of high-mean molecular weight species \citep{Brinkman2023B, 2021MNRAS.501.4148L, 2023arXiv230808687P}. There are potentially many more planets historically grouped with ``rocky planets'' due to their size, but which cannot have an accurate measurement of CMF due to the presence of a high-mean molecular weight envelope, such as GJ 3929 b, L 98-59 d, WASP-47 e, and 55 Cancri e \citep{2022ApJ...936...55B, 2021A&A...653A..41D, 2022AJ....163..197B, 2018A&A...619A...1B, 2024arXiv240504744H}. Phase-curve observations of many of these worlds with JWST would allow us to distinguish bare rocky planets from those with gaseous envelopes and demonstrate if these planets are genuinely low density, or if they too have an Earth-like interior composition.

\section{Acknowledgements}


The authors wish to recognize and acknowledge the very significant cultural role and reverence that the summit of Maunakea has always had within the Native Hawaiian community. We are most fortunate to have the opportunity to conduct observations from this mountain.

This work is based on observations obtained at the international Gemini Observatory, a program of NSF NOIRLab, which is managed by the Association of Universities for Research in Astronomy (AURA) under a cooperative agreement with the U.S. National Science Foundation on behalf of the Gemini Observatory partnership: the U.S. National Science Foundation (United States), National Research Council (Canada), Agencia Nacional de Investigaci\'{o}n y Desarrollo (Chile), Ministerio de Ciencia, Tecnolog\'{i}a e Innovaci\'{o}n (Argentina), Minist\'{e}rio da Ci\^{e}ncia, Tecnologia, Inova\c{c}\~{o}es e Comunica\c{c}\~{o}es (Brazil), and Korea Astronomy and Space Science Institute (Republic of Korea).

C.L.B. is supported by the National Science Foundation Graduate Research Fellowship under Grant No. 1842402 and NASA's Interdisciplinary Consortia for Astrobiology Research (NNH19ZDA001N-ICAR) under award number 19-ICAR19 2-0041. 

L.M.W. acknowledges support from the NASA Exoplanet Research Program through grant 80NSSC23K0269.

D.H. also acknowledges support from the Alfred P. Sloan Foundation.

C.L.B. and L.M.W. also acknowledge support from NASA Keck Key Stragetic Mission Support Grant No. 80NSSC19K1475.

N.S. acknowledges support by the National Science Foundation Graduate Research Fellowship Program under Grant Numbers 1842402 \& 2236415.

JMJO acknowledges support from NASA through the NASA Hubble Fellowship
grant HST-HF2-51517.001, awarded by STScI. STScI is operated by the
Association of Universities for Research in Astronomy, Incorporated,
under NASA contract NAS5-26555.

This research was carried out, in part, at the Jet Propulsion Laboratory and the California Institute of Technology under a contract with the National Aeronautics and Space Administration and funded through the President’s and Director’s Research \& Development Fund Program.

%
\facilities{Transiting Exoplanet Survey Satellite (TESS), W. M. Keck Observatory, Gemini Observatory}

\software{Radvel \citep{2018PASP..130d4504F}, SuperEarth, \citep{2006Icar..181..545V, Plotnykov2020}, 
         NumPy \citep{harris2020array}, Matplotlib \citep{Hunter:2007}, pandas \citep{mckinney-proc-scipy-2010}, Astropy \citep{astropy:2013, astropy:2018, astropy:2022}, SciPy \citep{2020SciPy-NMeth}   }
\pagebreak

\begin{center}

\section*{Appendix}

\end{center}

Radial Velocities for TOI-561 b can be found in \cite{Brinkman2023B}. The RVs for Kepler-10, Kepler-100, Kepler-407, TOI-1011, TOI-1444, and HD 93963 A can be found here. 

\begin{center}
    Kepler-10
\end{center}
\begin{longtable}{lccl}
\toprule
          Time &     RV &    RV error & Instrument \\
          $\rm [BJD - 2450000]$ & [m/s] & [m/s] & \\\midrule
9440.926139 &  -1.24 &    1.80 &  MAROON-X-Blue \\
9440.926139 &  -7.69 &    3.20 &   MAROON-X-Red \\
9441.749125 &  -6.24 &    1.18 &  MAROON-X-Blue \\
9441.749125 &  -1.29 &    2.19 &   MAROON-X-Red \\
9441.860056 &  -3.45 &    1.22 &  MAROON-X-Blue \\
9441.860056 &  -5.78 &    2.31 &   MAROON-X-Red \\
9441.868126 &  -4.71 &    1.59 &        HIRES \\
9442.011041 &  -7.71 &    1.59 &        HIRES \\
9443.822619 &   2.07 &    1.40 &        HIRES \\
9443.890404 &  -2.17 &    2.55 &   MAROON-X-Red \\
9443.890404 &  -1.43 &    1.39 &  MAROON-X-Blue \\
9444.834895 &   8.43 &    2.43 &   MAROON-X-Red \\
9444.834895 &   4.12 &    1.31 &  MAROON-X-Blue \\
9444.977016 &  -3.89 &    1.52 &        HIRES \\
9445.070891 &   8.27 &    2.65 &        HIRES \\
9445.737586 &   6.74 &    1.55 &  MAROON-X-Blue \\
9445.737586 &   5.33 &    2.81 &   MAROON-X-Red \\
9445.871900 &   2.38 &    1.39 &  MAROON-X-Blue \\
9445.871900 &   3.03 &    2.55 &   MAROON-X-Red \\
9446.734040 &   0.69 &    3.25 &   MAROON-X-Red \\
9446.734040 &   3.42 &    1.81 &  MAROON-X-Blue \\
9446.845743 &   7.65 &    1.61 &        HIRES \\
9447.809723 &   0.01 &    1.95 &  MAROON-X-Blue \\
9447.809723 &  -8.02 &    3.53 &   MAROON-X-Red \\
9447.929888 &   0.72 &    3.55 &   MAROON-X-Red \\
9447.929888 &   5.47 &    2.08 &  MAROON-X-Blue \\
9449.845796 &   4.96 &    1.47 &        HIRES \\
9449.951385 &  10.64 &    1.50 &        HIRES \\
9451.851473 &   6.43 &    1.67 &        HIRES \\
9451.955418 &   8.44 &    1.59 &        HIRES \\
10105.894233 &   0.00 &    2.45 &         KPF \\
10105.944887 &   0.37 &    2.49 &         KPF \\
10105.980114 &   0.25 &    2.64 &         KPF \\
10106.036962 &  -2.87 &    2.38 &         KPF \\
10106.085065 &  -5.79 &    2.30 &         KPF \\
10123.798785 &  -3.31 &    1.34 &         KPF \\
10123.828720 &  -6.28 &    1.29 &         KPF \\
10123.904347 &  -2.28 &    1.25 &         KPF \\
10123.949288 &  -1.99 &    1.25 &         KPF \\
10123.986391 &  -1.08 &    1.22 &         KPF \\
10156.787412 &  40.05 &    1.39 &         KPF \\
10156.845773 &  42.30 &    1.40 &         KPF \\
10156.900819 &  42.26 &    1.33 &         KPF \\
10156.955785 &  40.92 &    1.56 &         KPF \\
10156.993560 &  39.88 &    1.92 &         KPF \\
\bottomrule
\end{longtable}

\begin{center}
    Kepler-100
\end{center}
\begin{longtable}{lccl}
\toprule
          Time &     RV &    RV error & Instrument \\
          $\rm [BJD - 2450000]$ & [m/s] & [m/s] & \\\midrule
 9435.880812 &  2.94 &    1.49 &            HIRES \\
 9439.893041 & -5.11 &    3.35 &       MAROON-X-Red \\
 9439.893041 &  2.69 &    1.75 &      MAROON-X-Blue \\
 9441.025237 & -2.57 &    1.52 &            HIRES \\
 9441.786036 &  9.66 &    2.37 &       MAROON-X-Red \\
 9441.786036 &  3.22 &    1.20 &      MAROON-X-Blue \\
 9442.743883 & -3.74 &    2.61 &       MAROON-X-Red \\
 9442.743883 & -2.24 &    1.38 &      MAROON-X-Blue \\
 9443.857676 & -1.32 &    1.38 &            HIRES \\
 9444.868527 & -3.71 &    2.58 &       MAROON-X-Red \\
 9444.868527 & -3.44 &    1.30 &      MAROON-X-Blue \\
 9445.860715 & -4.45 &    3.36 &       MAROON-X-Red \\
 9445.860715 & -3.34 &    1.77 &      MAROON-X-Blue \\
 9446.062469 &  0.83 &    1.86 &            HIRES \\
 9446.767743 & -0.77 &    1.83 &      MAROON-X-Blue \\
 9446.767743 & -3.81 &    3.48 &       MAROON-X-Red \\
 9447.919542 &  5.91 &    2.16 &      MAROON-X-Blue \\
 9447.919542 & -1.29 &    3.91 &       MAROON-X-Red \\
 9449.786904 &  1.66 &    1.30 &      MAROON-X-Blue \\
 9449.786904 &  0.63 &    2.59 &       MAROON-X-Red \\
 9449.898803 & -2.76 &    1.36 &            HIRES \\
 9450.956780 &  0.95 &    1.67 &            HIRES \\
 9451.999340 &  1.64 &    1.45 &            HIRES \\
 9452.796037 &  2.43 &    1.41 &            HIRES \\
 9663.096543 & -4.16 &    2.90 &  MAROON-X-Red \\
 9663.096543 & -0.86 &    1.46 & MAROON-X-Blue \\
 9665.060330 & -2.01 &    1.21 & MAROON-X-Blue \\
 9665.060330 & -2.13 &    2.55 &  MAROON-X-Red \\
 9667.118592 &  2.96 &    3.20 &  MAROON-X-Red \\
 9667.118618 &  4.42 &    1.59 & MAROON-X-Blue \\
 9668.095467 &  5.86 &    2.99 & MAROON-X-Blue \\
 9668.095467 & 15.07 &    6.34 &  MAROON-X-Red \\
 9672.140845 &  2.17 &    1.76 & MAROON-X-Blue \\
 9672.140845 &  9.14 &    3.74 &  MAROON-X-Red \\
 9674.135865 &  0.02 &    2.91 & MAROON-X-Blue \\
 9674.135865 & -3.14 &    6.18 &  MAROON-X-Red \\
 9694.003345 & -3.19 &    1.72 & MAROON-X-Blue \\
 9694.003345 & -8.92 &    3.73 &  MAROON-X-Red \\
 9769.951177 & -2.28 &    1.31 & MAROON-X-Blue \\
 9769.951177 &  2.59 &    2.68 &  MAROON-X-Red \\
 9772.054005 &  0.31 &    1.02 & MAROON-X-Blue \\
 9772.054005 &  0.97 &    2.15 &  MAROON-X-Red \\
 9774.836183 &  2.49 &    2.72 &  MAROON-X-Red \\
 9774.836183 &  0.72 &    1.35 & MAROON-X-Blue \\
 9779.083207 & -0.88 &    2.41 & MAROON-X-Blue \\
 9779.083207 & -0.13 &    4.71 &  MAROON-X-Red \\
 9780.886065 & -0.80 &    1.04 & MAROON-X-Blue \\
 9780.886065 & -5.70 &    2.22 &  MAROON-X-Red \\
 9782.043842 & -1.91 &    2.72 &  MAROON-X-Red \\
 9782.043842 &  2.07 &    1.33 & MAROON-X-Blue \\
 9786.019063 &  2.69 &    1.16 & MAROON-X-Blue \\
 9786.019063 &  0.17 &    2.43 &  MAROON-X-Red \\
 9793.008317 &  2.45 &    3.10 &  MAROON-X-Red \\
 9793.008317 & -3.00 &    1.49 & MAROON-X-Blue \\
10118.941204 & -0.92 &    0.86 & MAROON-X-Blue \\
10118.941204 &  3.69 &    1.90 &  MAROON-X-Red \\
10120.082883 & -1.27 &    1.28 & MAROON-X-Blue \\
10120.082883 & -6.46 &    2.74 &  MAROON-X-Red \\
10120.889558 &  2.94 &    1.95 &  MAROON-X-Red \\
10120.889558 &  0.88 &    0.91 & MAROON-X-Blue \\
10122.942551 &  1.20 &    1.16 & MAROON-X-Blue \\
10122.942551 &  4.69 &    2.45 &  MAROON-X-Red \\
10124.037849 & -3.82 &    1.68 &  MAROON-X-Red \\
10124.037849 & -0.77 &    0.76 & MAROON-X-Blue \\
10124.942434 &  2.11 &    1.78 &  MAROON-X-Red \\
10124.942434 & -1.06 &    0.81 & MAROON-X-Blue \\
10130.994088 &  0.12 &    0.93 & MAROON-X-Blue \\
10130.994088 &  0.04 &    1.97 &  MAROON-X-Red \\
10131.902607 & -1.43 &    2.00 &  MAROON-X-Red \\
10131.902607 &  3.14 &    0.96 & MAROON-X-Blue \\
10133.932088 &  0.90 &    0.71 & MAROON-X-Blue \\
10133.932088 & -0.62 &    1.55 &  MAROON-X-Red \\
10136.039190 &  1.20 &    0.89 & MAROON-X-Blue \\
10136.039190 & -3.61 &    1.91 &  MAROON-X-Red \\
\bottomrule
\end{longtable}

\begin{center}
    Kepler-407
\end{center}
\begin{longtable}{lccl}\toprule
          Time &     RV &    RV error & Instrument \\
          $\rm [BJD - 2450000]$ & [m/s] & [m/s] & \\
\midrule
9440.944230 & 0.59 & 2.65 & MAROON-X-Red \\
9440.944230 & -2.01 & 1.80 & MAROON-X-Blue \\
9441.765978 & -4.10 & 1.73 & MAROON-X-Red \\
9441.765978 & -3.50 & 1.10 & MAROON-X-Blue \\
9441.854656 & -30.06 & 1.79 & HIRES \\
9441.876978 & -4.37 & 1.27 & MAROON-X-Blue \\
9441.876978 & 0.98 & 1.94 & MAROON-X-Red \\
9441.995685 & -29.19 & 1.98 & HIRES \\
9443.810943 & -26.87 & 1.64 & HIRES \\
9444.851442 & -6.79 & 1.80 & MAROON-X-Red \\
9444.851442 & -6.35 & 1.14 & MAROON-X-Blue \\
9444.936380 & -3.04 & 1.73 & MAROON-X-Red \\
9444.936380 & -2.54 & 1.19 & MAROON-X-Blue \\
9444.963707 & -40.69 & 1.63 & HIRES \\
9445.057757 & -36.88 & 1.93 & HIRES \\
9445.842010 & -1.09 & 2.42 & MAROON-X-Red \\
9445.842010 & 1.53 & 1.59 & MAROON-X-Blue \\
9445.925438 & -1.45 & 2.03 & MAROON-X-Red \\
9445.925438 & -1.10 & 1.41 & MAROON-X-Blue \\
9446.750100 & 2.66 & 2.52 & MAROON-X-Red \\
9446.750100 & 1.20 & 1.68 & MAROON-X-Blue \\
9446.828846 & -27.10 & 1.77 & HIRES \\
9447.827762 & 4.04 & 2.09 & MAROON-X-Blue \\
9447.827762 & -3.33 & 3.18 & MAROON-X-Red \\
9447.948669 & 4.61 & 2.10 & MAROON-X-Blue \\
9447.948669 & -0.72 & 2.88 & MAROON-X-Red \\
9449.740225 & 4.76 & 1.16 & MAROON-X-Blue \\
9449.740225 & 4.64 & 1.80 & MAROON-X-Red \\
9449.832914 & -26.77 & 1.71 & HIRES \\
9449.862948 & 4.83 & 1.15 & MAROON-X-Blue \\
9449.862948 & 4.10 & 1.79 & MAROON-X-Red \\
9449.938641 & -21.23 & 1.81 & HIRES \\
9449.945274 & 7.75 & 1.35 & MAROON-X-Blue \\
9449.945274 & 1.92 & 1.81 & MAROON-X-Red \\
9451.838659 & -28.81 & 1.77 & HIRES \\
9451.940081 & -27.10 & 1.87 & HIRES \\
10117.874729 & -11.62 & 2.42 & KPF \\
10117.927737 & -10.90 & 2.41 & KPF \\
10123.856261 & -1.44 & 1.48 & KPF \\
10123.892974 & -0.23 & 1.47 & KPF \\
10123.938194 & 0.00 & 1.45 & KPF \\
10123.997274 & -0.34 & 1.41 & KPF \\
10124.092360 & -1.97 & 1.54 & KPF \\
10156.775616 & 19.02 & 1.59 & KPF \\
10156.834540 & 21.25 & 1.60 & KPF \\
10156.887944 & 21.13 & 1.65 & KPF \\
10156.944165 & 20.58 & 1.74 & KPF \\
10156.981183 & 23.17 & 1.83 & KPF \\
10157.006553 & 18.50 & 2.13 & KPF \\
\bottomrule
\end{longtable}\par

\begin{center}
    TOI-1011
\end{center}
\begin{longtable}{lccl}
\toprule
          Time &     RV &    RV error & Instrument \\
          $\rm [BJD - 2450000]$ & [m/s] & [m/s] & \\\midrule
 9663.724251 &  -1.88 &    0.69 &      MAROON-X Blue \\
 9663.799574 &  -1.87 &    0.58 &      MAROON-X Blue \\
 9664.725262 &   0.64 &    0.86 &      MAROON-X Blue \\
 9664.808405 &  -0.22 &    0.77 &      MAROON-X Blue \\
 9665.723737 &   0.56 &    0.54 &      MAROON-X Blue \\
 9665.781508 &  -1.68 &    0.65 &      MAROON-X Blue \\
 9666.716583 &  -3.07 &    0.57 &      MAROON-X Blue \\
 9666.764492 &  -5.19 &    0.49 &      MAROON-X Blue \\
9670.717911 &   1.10 &    0.73 &      MAROON-X Blue \\
9671.713411 &  -0.33 &    0.73 &      MAROON-X Blue \\
9671.760033 &  -2.33 &    0.91 &      MAROON-X Blue \\
9672.721435 &   4.19 &    0.62 &      MAROON-X Blue \\
 9677.788349 &  -0.31 &    0.83 &      MAROON-X Blue \\
9680.726256 &  -0.98 &    0.66 &      MAROON-X Blue \\
9680.766194 &  -2.85 &    0.66 &      MAROON-X Blue \\
9682.725764 &   1.62 &    0.62 &      MAROON-X Blue \\
9682.766791 &   1.78 &    0.66 &      MAROON-X Blue \\
9683.727533 &  -1.50 &    0.57 &      MAROON-X Blue \\
9684.732398 &   1.25 &    0.72 &      MAROON-X Blue \\
9684.736867 &   3.52 &    0.57 &      MAROON-X Blue \\
9684.770912 &   2.14 &    0.56 &      MAROON-X Blue \\
9685.754961 &  -0.01 &    0.76 &      MAROON-X Blue \\
9689.732128 &   3.33 &    0.45 &      MAROON-X Blue \\
 9663.724251 &   1.40 &    1.20 &       MAROON-X Red \\
 9663.799574 &  -0.77 &    1.01 &       MAROON-X Red \\
 9664.725262 &  -0.35 &    1.38 &       MAROON-X Red \\
 9664.808405 &   0.72 &    1.32 &       MAROON-X Red \\
 9665.723737 &   0.96 &    0.95 &       MAROON-X Red \\
 9665.781508 &   2.45 &    1.13 &       MAROON-X Red \\
 9666.716583 &  -2.44 &    1.00 &       MAROON-X Red \\
 9666.764492 &  -2.96 &    0.88 &       MAROON-X Red \\
9670.717911 &   2.33 &    1.24 &       MAROON-X Red \\
9671.713411 &  -1.64 &    1.24 &       MAROON-X Red \\
 9671.760033 &   0.40 &    1.54 &       MAROON-X Red \\
9672.721435 &   3.53 &    1.09 &       MAROON-X Red \\
9677.788349 &  -2.15 &    1.39 &       MAROON-X Red \\
9680.726256 &  -3.32 &    1.12 &       MAROON-X Red \\
9680.766194 &  -4.45 &    1.12 &       MAROON-X Red \\
9682.725764 &   0.37 &    1.07 &       MAROON-X Red \\
9682.766791 &   0.14 &    1.14 &       MAROON-X Red \\
9683.727533 &  -2.32 &    0.99 &       MAROON-X Red \\
9684.732398 &   1.09 &    1.12 &       MAROON-X Red \\
9684.736867 &   1.85 &    0.99 &       MAROON-X Red \\
9684.770912 &   0.13 &    0.97 &       MAROON-X Red \\
9685.754961 &   1.45 &    1.30 &       MAROON-X Red \\
9689.732128 &   2.55 &    0.84 &       MAROON-X Red \\
10284.022997 &   2.93 &    0.42 &  MAROON-X Blue \\
10284.088992 &   3.75 &    0.39 &  MAROON-X Blue \\
10284.976994 &   4.44 &    0.41 &  MAROON-X Blue \\
 10285.942445 &   2.90 &    0.37 &  MAROON-X Blue \\
10286.142139 &   4.02 &    0.57 &  MAROON-X Blue \\
10286.977494 &   0.09 &    0.47 &  MAROON-X Blue \\
10287.026383 &  -1.06 &    0.45 &  MAROON-X Blue \\
10287.090683 &   1.26 &    0.39 &  MAROON-X Blue \\
10287.973298 &   2.27 &    0.65 &  MAROON-X Blue \\
10288.034753 &   3.51 &    0.49 &  MAROON-X Blue \\
10288.095746 &   0.88 &    0.67 &  MAROON-X Blue \\
10289.959604 &   0.25 &    0.51 &  MAROON-X Blue \\
10290.134294 &   0.49 &    0.55 &  MAROON-X Blue \\
10290.952287 &  -0.80 &    0.39 &  MAROON-X Blue \\
10291.010149 &  -2.27 &    0.42 &  MAROON-X Blue \\
10292.105021 &  -2.42 &    0.48 &  MAROON-X Blue \\
10293.014414 &   0.01 &    1.03 &  MAROON-X Blue \\
10293.063049 &   2.03 &    0.85 &  MAROON-X Blue \\
10294.019021 &  -2.93 &    0.59 &  MAROON-X Blue \\
10294.079182 &  -2.81 &    0.59 &  MAROON-X Blue \\
10295.942686 &  -0.64 &    0.49 &  MAROON-X Blue \\
10296.020741 &  -1.30 &    0.56 &  MAROON-X Blue \\
10296.079124 &  -2.41 &    0.53 &  MAROON-X Blue \\
10297.099274 &  -2.29 &    0.46 &  MAROON-X Blue \\
10301.908245 &  -7.56 &    0.53 &  MAROON-X Blue \\
10301.960064 &  -6.06 &    0.62 &  MAROON-X Blue \\
10303.077615 &  -5.96 &    0.66 &  MAROON-X Blue \\
10284.022997 &   0.45 &    0.77 &   MAROON-X Red \\
10284.088992 &   3.19 &    0.72 &   MAROON-X Red \\
10284.976994 &   5.44 &    0.75 &   MAROON-X Red \\
10285.942445 &   3.52 &    0.69 &   MAROON-X Red \\
10286.142139 &   1.56 &    1.00 &   MAROON-X Red \\
10286.977494 &  -0.66 &    0.85 &   MAROON-X Red \\
10287.026383 &  -2.12 &    0.81 &   MAROON-X Red \\
10287.090683 &  -0.43 &    0.72 &   MAROON-X Red \\
10287.973298 &   2.09 &    1.17 &   MAROON-X Red \\
10288.034753 &   3.23 &    0.88 &   MAROON-X Red \\
10288.095746 &   5.76 &    1.19 &   MAROON-X Red \\
10289.959604 &  -1.86 &    0.91 &   MAROON-X Red \\
10290.134294 &   0.03 &    0.98 &   MAROON-X Red \\
10290.952287 &   0.37 &    0.72 &   MAROON-X Red \\
10291.010149 &  -2.47 &    0.78 &   MAROON-X Red \\
10292.105021 &  -3.15 &    0.88 &   MAROON-X Red \\
10293.014414 &   2.48 &    1.88 &   MAROON-X Red \\
10293.063049 &   2.34 &    1.50 &   MAROON-X Red \\
10294.019021 &  -2.42 &    1.08 &   MAROON-X Red \\
10294.079182 &  -3.98 &    1.06 &   MAROON-X Red \\
10295.942686 &  -1.19 &    0.91 &   MAROON-X Red \\
10296.020741 &  -3.43 &    1.02 &   MAROON-X Red \\
10296.079124 &  -2.82 &    0.97 &   MAROON-X Red \\
10297.099274 &  -2.10 &    0.88 &   MAROON-X Red \\
10301.908245 &  -5.44 &    0.98 &   MAROON-X Red \\
10301.960064 &  -6.05 &    1.13 &   MAROON-X Red \\
10303.077615 &  -5.61 &    1.20 &   MAROON-X Red \\
\bottomrule
\end{longtable}

\begin{center}
    TOI-1444
\end{center}
\begin{longtable}{lccl}
\toprule
          Time &     RV &    RV error & Instrument \\
          $\rm [BJD - 2450000]$ & [m/s] & [m/s] & \\\midrule
  10176.772964 & -0.40 &  1.78 &  KPF \\
  10176.826928 &  9.31 &  1.71 &  KPF \\
  10176.855716 &  7.18 &  1.61 &  KPF \\
  10176.887327 &  6.53 &  1.81 &  KPF \\
  10176.926072 &  7.59 &  1.53 &  KPF \\
  10182.782029 &  5.15 &  1.64 &  KPF \\
  10182.829524 &  5.18 &  1.60 &  KPF \\
  10182.933259 &  1.81 &  1.67 &  KPF \\
  10182.973064 & -0.47 &  2.05 &  KPF \\
  10183.756424 &  1.06 &  1.59 &  KPF \\
  10183.798456 & -2.04 &  1.68 &  KPF \\
  10183.852255 & -3.65 &  1.51 &  KPF \\
  10183.928530 & -2.60 &  1.62 &  KPF \\
  10183.974074 & -1.32 &  1.87 &  KPF \\
  10184.745364 &  0.08 &  1.71 &  KPF \\
  10184.807807 & -1.34 &  1.62 &  KPF \\
  10184.856484 & -2.72 &  1.83 &  KPF \\
  10184.877824 & -2.76 &  1.59 &  KPF \\
  10184.924067 &  0.00 &  1.63 &  KPF \\\bottomrule
\end{longtable}\par

\begin{center}
    HD 94963 A
\end{center}
\begin{longtable}{lccl}
\toprule
Time &     RV &    RV error & Instrument \\
          $\rm [BJD - 2450000]$ & [m/s] & [m/s] & \\\midrule
10253.094570 &  14.44 &    2.58 &  KPF \\
10253.132488 &  11.22 &    2.45 &  KPF \\
10259.122080 &   3.09 &    2.70 &  KPF \\
10259.139811 &   8.03 &    2.80 &  KPF \\
10259.153816 &   6.04 &    2.76 &  KPF \\
10269.063718 &  17.08 &    3.19 &  KPF \\
10269.106397 &  18.98 &    2.85 &  KPF \\
10269.155494 &  20.83 &    2.44 &  KPF \\
10273.058158 &   1.48 &    2.72 &  KPF \\
10273.079978 &   0.14 &    2.74 &  KPF \\
10273.124516 &   5.69 &    2.60 &  KPF \\
10274.081316 &  -1.78 &    2.55 &  KPF \\
10274.097197 &  -3.59 &    2.24 &  KPF \\
10274.128299 &  -0.68 &    2.22 &  KPF \\
10274.141901 &  -4.27 &    2.16 &  KPF \\
10308.110805 & -17.32 &    3.38 &  KPF \\
10308.149269 & -13.51 &    2.77 &  KPF \\
10308.985576 &  -3.10 &    2.79 &  KPF \\
10309.065193 &  -6.92 &    2.63 &  KPF \\
10309.088836 &  -2.97 &    2.46 &  KPF \\
10309.109569 &  -6.08 &    2.45 &  KPF \\
10309.175918 &  -5.69 &    2.32 &  KPF \\
10330.014003 &  -2.00 &    2.52 &  KPF \\
10330.051287 &   2.28 &    2.95 &  KPF \\
10330.079980 &  -0.94 &    2.53 &  KPF \\
10330.088131 &   0.00 &    2.62 &  KPF \\
10330.130293 &   1.13 &    2.51 &  KPF \\
\bottomrule
\end{longtable}




\bibliography{paper}{}
\bibliographystyle{aasjournal}



\end{document}